\documentclass[journal,twoside]{IEEEtran}
\IEEEoverridecommandlockouts
\usepackage{cite}

\ifCLASSINFOpdf
  \usepackage[pdftex]{graphicx}
  \graphicspath{{../images/}}
  \DeclareGraphicsExtensions{.pdf,.jpeg,.png}
\else
  \usepackage[dvips]{graphicx}
  \graphicspath{{../images/}}
  \DeclareGraphicsExtensions{.eps}
\fi
\usepackage[cmex10]{amsmath}
\usepackage{amssymb}
\ifCLASSOPTIONcompsoc
  \usepackage[caption=false,font=normalsize,labelfont=sf,textfont=sf]{subfig}
\else
  \usepackage[caption=false,font=footnotesize]{subfig}
\fi
\usepackage{tikz}

\usepackage{mathrsfs}
\usepackage{multirow}

\usepackage{acronym}

\usepackage{pgfplots} 
\pgfplotsset{compat=newest} 
\pgfplotsset{plot coordinates/math parser=false} 
\newlength\figureheight 
\newlength\figurewidth 
\usepackage{tikzscale}
\usepackage{epstopdf}

\usepackage{cleveref}

\newcommand{\conv}[2]{\left\langle #1 , #2 \right\rangle}

\newtheorem{theorem}{Theorem}

\newtheorem{rem}{Remark}

\newcommand{\setF}{\boldsymbol{X}}
\newcommand{\setU}{\boldsymbol{U}}
\newcommand{\setX}{\boldsymbol{X}}
\newcommand{\setY}{\boldsymbol{Y}}
\newcommand{\statef}{\boldsymbol{x}}
\newcommand{\statex}{\boldsymbol{s}}
\newcommand{\statexintro}{\boldsymbol{x}}
\newcommand{\statey}{{\boldsymbol{z}_\mathrm{G}}}

\newcommand{\setZ}{\boldsymbol{Z}}

\newcommand{\setZFA}{\boldsymbol{Z}^\mathrm{FA}}
\newcommand{\setZD}{\boldsymbol{Z}^\mathrm{D}}
\newcommand{\statez}{\boldsymbol{z}}
\newcommand{\diff}{\mathrm{d}}

\newcommand{\sigmaG}{\sigma_{\mathrm{G}}^{2}}
\newcommand{\sigmaVtoF}{\sigma_{\mathrm{V2F}}^{2}}

\let\oldFootnote\footnote
\newcommand\nextToken\relax
\renewcommand\footnote[1]{%
	\oldFootnote{#1}\futurelet\nextToken\isFootnote}
\newcommand\isFootnote{%
	\ifx\footnote\nextToken\textsuperscript{,}\fi}

\usepackage[normalem]{ulem}

\acrodef{ad}[AD]{autonomous driving}
\acrodef{cdf}[CDF]{cumulative distribution function}
\acrodef{cv}[CV]{constant velocity}
\acrodef{da}[DA]{data association}
\acrodef{dglmb}[$\delta$-GLMB]{$\delta$-generalized labeled multi-Bernoulli}
\acrodef{ekf}[EKF]{extended Kalman filter}
\acrodef{fisst}[FISST]{finite-set statistics}
\acrodef{fg}[FG]{factor graph}
\acrodef{fov}[FoV]{field-of-view}
\acrodef{gm}[GM]{Gaussian mixture}
\acrodef{gnss}[GNSS]{global navigation satellite system}
\acrodef{jpda}[JPDA]{joint probability data association}
\acrodef{kf}[KF]{Kalman filter}
\acrodef{kld}[KLD]{Kullback-Leibler divergence}
\acrodef{ldm}[LDM]{local dynamic map}
\acrodef{iid}[IID]{independent identically distributed}
\acrodef{imu}[IMU]{inertial measurment unit}
\acrodef{its}[ITS]{Intelligent transportation system}
\acrodef{v2v}[V2V]{vehicle-to-vehicle}
\acrodef{v2i}[V2I]{vehicle-to-infrastructure}
\acrodef{mb}[MB]{multi-Bernoulli}
\acrodef{mbm}[MBM]{multi-Bernoulli mixture}
\acrodef{mc}[MC]{Monte-Carlo}
\acrodef{mtt}[MTT]{multitarget tracking}
\acrodef{mht}[MHT]{multi-hypothesis tracking}
\acrodef{rb}[RB]{Rao-Blackwellization}
\acrodef{rmse}[RMSE]{root mean square error}
\acrodef{rfs}[RFS]{random finite set}
\acrodef{rsu}[RSU]{road side unit}
\acrodef{rv}[RV]{random vector}
\acrodef{slam}[SLAM]{simultaneous localization and mapping}
\acrodef{slat}[SLAT]{simultaneous localization and tracking}
\acrodef{tombp}[TOMB/P]{track-oriented marginal multiple target multi-Bernoulli/Poisson}
\acrodef{pmb}[PMB]{Poisson multi-Bernoulli}
\acrodef{pmbm}[PMBM]{Poisson multi-Bernoulli mixture}
\acrodef{pdf}[PDF]{probability density function}
\acrodef{pf}[PF]{particle filter}
\acrodef{pgfl}[p.g.fl.]{probability generating functional}
\acrodef{ppp}[PPP]{Poisson point process}
\acrodef{ukf}[UKF]{unscented Kalman filter}
\acrodef{v2f}[V2F]{vehicle-to-feature}
\acrodef{ospa}[OSPA]{optimal sub-pattern assignment}
\acrodef{phd}[PHD]{probability hypothesis density}
\acrodef{wrt}[w.r.t.]{with respect to}
\acrodef{v2x}[V2X]{V2X}
\acrodef{gps}[GPS]{Global Positioning System}
\acrodef{rtkgps}[RTK]{Real Time Kinematic}
\acrodef{sbasgps}[SBAS]{Satellite-Based Augmentation System}
\acrodef{spsgps}[SPS]{Standard Positioning System}
\acrodef{dgnss}[DGPS]{Differential Global Positioning System}

\newtheorem{lemma}[theorem]{Lemma}

\newcommand{\approptoinn}[2]{\mathrel{\vcenter{
			\offinterlineskip\halign{\hfil$##$\cr
				#1\propto\cr\noalign{\kern2pt}#1\sim\cr\noalign{\kern-2pt}}}}}
\newcommand{\appropto}{\mathpalette\approptoinn\relax}

\begin{document}
\tikzstyle{state}=[shape=circle,draw=blue!90,fill=blue!10,line width=1pt]
%
\title{Multisensor Poisson Multi-Bernoulli Filter for Joint Target-Sensor State Tracking}

\author{Markus~Fr\"ohle,
Christopher~Lindberg,
Karl~Granstr\"om, 
Henk~Wymeersch 

\thanks{M. Fr\"ohle, K. Granstr\"om, and H. Wymeersch are with the Department of Electrical Engineering, Chalmers University of Technology, Gothenburg, Sweden. 
	C. Lindberg is with DENSO Sweden AB, Gothenburg, Sweden. E-mail: \texttt{ \{frohle, karl.granstrom, henkw, chrlin\}@chalmers.se}.
	This work was supported, in part, by the EU-H2020 project HIGHTS (High Precision Positioning for Cooperative ITS Applications) under grant no.~MG-3.5a-2014-636537 and COPPLAR (campus shuttle cooperative perception and planning platform) project, funded under grant no. 2015-04849 from Vinnova.
}
}

\maketitle

\begin{abstract}

In a typical \ac{mtt} scenario, the sensor state is either assumed known, or tracking is performed in the sensor's (relative) coordinate frame. This assumption does not hold when the sensor, e.g., an automotive radar, is mounted on a vehicle, and the target state should be represented in a global (absolute) coordinate frame. Then it is important to consider the uncertain location of the vehicle on which the sensor is mounted for \ac{mtt}. 
In this paper, we present a multisensor low complexity Poisson multi-Bernoulli \ac{mtt} filter, which jointly tracks the uncertain vehicle state and target states. Measurements collected by different sensors mounted on multiple vehicles with varying location uncertainty are incorporated sequentially based on the arrival of new sensor measurements. In doing so, targets observed from a sensor mounted on a well-localized vehicle reduce the state uncertainty of other poorly localized vehicles, provided that a common non-empty subset of targets is observed. {A low complexity filter is obtained by approximations of the joint sensor-feature state density minimizing the \ac{kld}.}
Results from synthetic as well as experimental measurement data, collected in a vehicle driving scenario, demonstrate the performance benefits of joint vehicle-target state tracking. 

\end{abstract}


%
\IEEEpeerreviewmaketitle

%

\section{Introduction}
\label{sec:Intro}\label{sec:motivation}
\acp{its} in general, and \ac{ad} in particular, require accurate position information \cite{leonard2008perception}. Measurements provided by various on-board sensors allow to infer the vehicle state, e.g., position and velocity, as well as information about the surrounding environment. For instance, a \ac{gnss} receiver provides absolute position, whereas a radar sensor provides relative position of a feature \ac{wrt} the sensor origin. Furthermore, vehicles have access to a pre-recorded \ac{ldm} containing static features such as, e.g., landmarks \cite{cadena2016past}. 
Dynamic features such as pedestrians, cyclists, etc., are not part of the pre-recorded map. For an \ac{ad} system to be fully aware of the surrounding environment, dynamic features need to be estimated from measurements and tracked over time, using the vehicle's on-board sensors, thus allowing to enrich the vehicle's \ac{ldm}. 

In order to update the \ac{ldm} by new features (static, dynamic) described in a global coordinate frame, location uncertainty of the platform, where the sensors are mounted (e.g., a vehicle), needs to be considered. Furthermore, information from one vehicle can be utilized to increase location accuracy of other vehicles, and vice versa \cite{meyer2016journal, soatti2017conf}.
In the literature, three different research tracks can be discerned: (i) \ac{mtt} when both features and sensors are mobile, but the sensor states are known; (ii) \ac{slam} when the sensor state and feature state are unknown, but features are static; (iii) \ac{slat} combines \ac{mtt} and \ac{slam} by considering unknown mobile feature and mobile sensor states. All three tracks include measurements due to clutter; missed detections; unknown measurement-to-target correspondence; and target appearance and disappearance.

MTT filters based on \ac{rfs} and \ac{fisst}, see, e.g., \cite{Mahler2007,mahler2014advances} for recent works and \cite{mori1984multitarget, mori1986tracking} for earlier works, have gained much attention. For example, the \ac{phd} filter propagates the first moment of the \ac{rfs} density over time \cite{vo2006gaussian, vo2005smc, vo2007analytic}.
The \ac{pmbm} filter models unknown, i.e., never detected, features by a Poisson process and detected features by a \ac{mbm} \cite{williams2015marginal}. Based on this, the \ac{tombp} filter approximates the global joint \ac{da} by the product of marginal \acp{da}, similar to the \ac{jpda} filter, which allows a computationally efficient implementation \cite{williams2015marginal}.
In \cite{garcia2017poisson}, a derivation of the \ac{pmbm} filter based on standard single target measurement models, without using \acp{pgfl} and functional derivatives, is presented. Furthermore, connections to the \ac{dglmb} filter \cite{vo2014labeled,reuter2014labeled} are discussed.
In contrast to the above \ac{fisst} approaches, \cite{meyer2018,meyer2017scalable} proposed a \ac{fg} based approach for a variant of the \ac{jpda} filter. A multi-scan scenario was considered, and the filter was realized by running loopy belief propagation on the \ac{fg} containing cycles.
Additionally, similarities between the \ac{fg} approach and the \ac{tombp} filter are discussed.

For static features (called landmarks) observed by a sensor with unknown state, \ac{slam} based methods can be employed to jointly estimate the sensor state and the landmarks (see, e.g., \cite{durrant2006simultaneous,cadena2016past}). %
In \cite{mullane2011random} and \cite{brekke2014novel}, an \ac{rfs} based approach to the \ac{slam} problem was proposed. The landmark state is conditioned on the sensor location and then tracked through a \ac{phd} filter following a \ac{rb}.

Joint estimation of the unknown sensor and mobile feature states, also termed \ac{slat}, was for instance addressed in \cite{lee2013slam}, where a combination of static and mobile features are estimated through \ac{mtt}.
In doing so, a particle based representation of the sensor state is combined with a parametric formulation of the \ac{phd} filter. 
Related to this, \cite{uney2016distributed, uney2016cooperative} considers the problem of sensor state estimation through target tracking in the \ac{rfs} framework by combining local \ac{phd} filters with the help of belief propagation. 
To achieve distributed processing, approximations in terms of separable likelihoods are taken. 
In \cite{teng2012distributed}, a particle based \ac{mtt} filter is presented for \ac{slat} in wireless sensor networks. Here, the measurement-to-target \ac{da} is known, but corrupted by noise leading to false alarms. In \cite{kantas2012distributed}, a message passing based distributed multisensor \ac{mtt} filter modeling target and sensor states by Gaussian \acp{pdf} is presented. Measurement-to-target DA is known and targets are always present and no false alarm measurements occur.  Similar to \cite{meyer2017scalable}, 
 a \ac{fg} based approach was considered in \cite{frohle2018wcnc} for an urban \ac{its} scenario, where the number of features is assumed a priori known, and in \cite{soatti2017conf} where the \ac{da} assumed to be known as well. %
In this paper, we consider the problem of multisensor \ac{slat} for joint estimation of the unknown sensor and feature states,
enabling accurate feature tracking \ac{wrt} the sensor uncertainty. 
Our proposed \ac{mtt} filter builds upon the Bayesian \ac{rfs} based \ac{pmbm} filter \cite{williams2015marginal} with proper feature birth and death processes, but explicitly models the sensor state uncertainty. {A low complexity filter is obtained by approximations of the joint sensor-feature state density minimizing the \ac{kld}. A tractable implementation is achieved through approximations similar to the \ac{tombp} filter \cite{williams2015marginal}.} The proposed \ac{mtt} filter allows to track the state of the sensor platform not only by (direct) measurements of the platform itself, but also through feature tracking in a multi-sensor setup.
The main contributions are:
\begin{itemize}
	\item A low-complexity asynchronous multisensor \ac{mtt} filter with uncertain sensor state information, 
	\item Sensor state tracking by fusion of multisensor \ac{mtt} information with local sensor information,
	\item Validation of the filter with real sensor data, as well as in-depth analysis of performance 
	with synthetic data. 
\end{itemize}

The remainder of this paper is organized as follows; Section~\ref{sec:backgroundRFS}
gives some background knowledge on \ac{rfs}, and Section~\ref{sec:ProblemFormulationAndSystemModels} introduces the problem
formulation and system models. Section~\ref{sec:MBfiltering} details the proposed \ac{mtt} filter with uncertain sensor state, including the %
 multisensor generalization of the proposed \ac{mtt} filter. Results with synthetic data are given in Section~\ref{sec:results-synthetic} and with experimental data in Section~\ref{sec:results-experimental}, respectively. Conclusions
are drawn in Section~\ref{sec:conclusions}.

\subsection*{Notation}
\label{sec:notation}
Scalars are described by non-bold letters $r$, vectors by lower-case bold letters $\statexintro$; matrices and sets by upper-case bold letters $\setX$. The cardinality of set $\setX$ is denoted $\vert \setX \vert$.
The set operator $\uplus$ denotes the disjoint set union, i.e., $\setF^u \uplus \setF^d = \setF$ means $\setF^u \cup \setF^d = \setF$ and $\setF^u \cap \setF^d = \emptyset$. The vehicle/sensor state is reserved by letter $\statex$, the feature state by letter $\statef$, and measurements by letter $\statez$. The identity matrix of size $n\times n$ is denoted $\boldsymbol{I}_n$. The $\ell_2$-norm of vector $\statexintro$ is $\Vert\statexintro\Vert_2$. 




\section{Background on \ac{rfs}}
\label{sec:backgroundRFS}
In this section, we describe some useful properties of an \ac{rfs}. If not stated otherwise, the source of all these is \cite{williams2015marginal}.
\subsection{Random Finite Set Formulation}
According to \cite{williams2015marginal}, \ac{rfs} based
 methods have been developed in \cite{Mahler2007} to conduct statistical
inference in problems in which the variables of interest and/or observations
form finite sets. In tracking, they address two major challenges of
interest: (i) the number of targets present in the scene is unknown,
(ii) measurements are invariant to ordering (measurement-to-target correspondence
is unknown). An \ac{rfs} $\setX$ is a finite-set valued random variable, which
can be described by a discrete probability distribution $p(n),$ where $n\geq0$ denotes the number of elements $\statexintro_i \in \setX$ for $i=0,\ldots,n$ 
and a family of joint \acp{pdf} $p_{n}(\statexintro_{1},\ldots,\statexintro_{n})$
yielding \cite{williams2015marginal} 
\begin{align}
f(\{\statexintro_{1},\ldots,\statexintro_{n}\})=p(n)\sum_{\pi}p_{n}(\statexintro_{\pi(1)},\ldots,\statexintro_{\pi(n)}),
\end{align}
where the sum spans over the $n!$ permutation functions $\pi(\cdot)$, such that
its \ac{rfs} density $f(\setX$) is permutation invariant. The set integral of a real-valued function $g(\setX)$ of a finite-set variable $\setX$ is defined as \cite[p. 361]{Mahler2007}, \cite{williams2015marginal}
\begin{align}
\int g(\setX)\delta \setX\triangleq g(\emptyset)+\sum_{n=1}^{\infty}\frac{1}{n!}\int g(\{\statexintro_{1},\ldots,\statexintro_{n}\})\diff\statexintro_{1}\cdots \diff\statexintro_{n}.
\end{align}

Two important examples of \ac{rfs}s are Bernoulli processes (and their generalization, \ac{mb} processes) and Poisson processes.  
A Bernoulli process $\setX$ with probability of existence $r$ and existence-conditioned
\ac{pdf} $p(\statexintro)$ has \ac{rfs} density %
\begin{equation}
f(\setX)=\begin{cases}
1-r  ,&\text{if }\setX=\emptyset,\\
r\cdot p(\statexintro)  ,&\text{if }\setX={\statexintro},\\
0  ,&\text{otherwise}.
\end{cases}\label{eq:BernoulliProcessDensity}
\end{equation}
The \ac{rfs} density of a \acf{mb} process $\setX$ can be expressed as
\begin{align}
f(\setX) = \sum_{\uplus_{i\in\mathbb{I}}\setX_i=\setX} \prod_{i\in\mathbb{I}}f_{i}(\setX_i)
\end{align}
for $\vert\setX\vert \leq \vert \mathbb{I} \vert$, and $f(\setX) =0$ otherwise. Here, $\mathbb{I}$  is the index set of the \ac{mb} with components $\{r_i,p_i(\statexintro)\}_{i\in\mathbb{I}}$.
A \ac{mbm} is a linear combination of \acp{mb} with density expressed by the Bernoulli components
\begin{align}
f(\setX)=\sum_{j\in\mathbb{J}}w_j f_j(\setX),
\end{align}
where $w_j$ is the weight of the $j$-th \ac{mb} with density $f_j(\setX)$ (such that $\sum_j w_j=1$), and $\mathbb{J}$ is the index set of the \ac{mbm}.
A \ac{ppp} with intensity function $\lambda_c(\boldsymbol{y})$
has \ac{rfs} density \cite{williams2015marginal}
\begin{align}
f(\setY)=e^{-\conv{\lambda_c}{1}}\prod_{\boldsymbol{y}\in \setY}\lambda_c(\boldsymbol{y})
\label{eq:backgroundPPP}
\end{align}
with inner product $\langle\lambda_c,h\rangle\triangleq\int\lambda_c(\boldsymbol{y})h(\boldsymbol{y})\diff\boldsymbol{y}.$
\begin{rem}
	If $\setX$ and $\setY$ are independent \ac{rfs}s such that $\boldsymbol{Z}=\setX\uplus \setY,$ then 
	\begin{align}
	f_{\boldsymbol{Z}}(\boldsymbol{Z}) 
	& = \sum_{\setX \uplus \setY = \boldsymbol{Z}} f_{\setX}(\setX) f_{\setY}(\setY).
	\label{eq:PMBMremark}
	\end{align}
	Note, when $\setX$ follows an \ac{mbm} process and $\setY$ an \ac{ppp} \eqref{eq:PMBMremark} is called a \ac{pmbm} density and \ac{pmb} density for $|\mathbb{J}|=1$.
\end{rem}

\begin{rem}
\label{sec:background-RFSstateEstimation}
A common way to estimate the set states from a Bernoulli
process with \ac{rfs} density $f(\setX)$ is by comparing the probability
of existence $r$ to an existence threshold $r_{\mathrm{th}}$. For $r>r_{\mathrm{th}},$ the target is
said to exist and has \ac{pdf} $p(\statexintro)$ (c.f. (\ref{eq:BernoulliProcessDensity})). Its state can then be estimated by the mean of $p(\statexintro)$, i.e., $\hat{\statexintro} = \int \statexintro p(\statexintro)\diff \statexintro$. See \cite[Sec.~VI]{garcia2017poisson}, for an extended discussion on state estimation.
\end{rem}

\subsection{Bayesian Filter Formulation}
\label{sec:bayesianFilterFormulation}
Similar to the \ac{rv} case, an \ac{rfs} based filter can be described, conceptually at least, within the Bayesian framework with alternating prediction and update steps operating on the state \ac{rfs} $\setX$ with \cite[Ch. 14]{Mahler2007}
\begin{align}
f_+(\setX)&=\int f(\setX|\setX')f_-(\setX')\delta \setX'
\label{eq:RFSfilterPrediction}
\end{align}
and 
\begin{align}
f(\setX|\boldsymbol{Z})\propto \ell(\boldsymbol{Z}|\setX)f_+(\setX),
\label{eq:BayesianUpdatePropTo}
\end{align}
where $f_-(\setX')$ is the prior \ac{rfs} density, $f(\setX|\setX')$
is the \ac{rfs} transition density, $f_+(\setX)$ is the predicted \ac{rfs} density, and $\ell(\boldsymbol{Z}|\setX)$ is the \ac{rfs} measurement likelihood for measurement set $\boldsymbol{Z}$. %

\subsection{Two Useful Lemmas}
\begin{lemma}\label{sec:lemma1}
A joint density $f(\setX,\setY)$ is approximated in the minimum \ac{kld} sense by \cite[Ch.~10]{bishop2006pattern}
\begin{align}
f(\setX,\setY) \approx \int f(\setX,\setY) \delta\setX \int f(\setX,\setY) \delta\setY.
\end{align}
\end{lemma}

\begin{lemma}\label{sec:lemma2}
The set integral can be expressed as \cite[Lemma~5]{williams2011graphical}
\begin{align}
\int \sum_{\setX \uplus \setY = \setF} f(\setX) g(\setY) \delta\setF 
= \int f(\setX)\delta\setX \int g(\setY)\delta\setY.
\end{align}
\end{lemma}

\section{Motivation, Problem Formulation and System Models}
\label{sec:ProblemFormulationAndSystemModels}

Here, we first present the motivation and problem formulation, as well as the vehicle and
feature dynamics. This is followed by the \ac{gnss} and \ac{v2f}
measurement models, and the communication assumption.
\subsection{Motivation}
\label{sec:motivation}
\begin{figure}[t]
	\centering
	\footnotesize
	\includegraphics{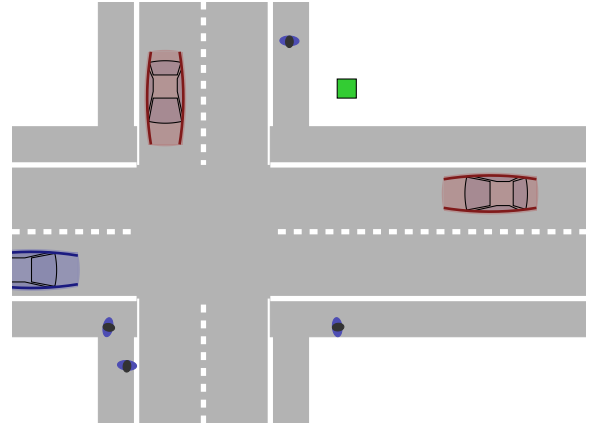}
	\caption{Urban ITS scenario with two vehicles (red color) cooperating through the \ac{rsu} (green color) and five mobile features (blue color).\label{fig:scenario}}
\end{figure}

We consider an urban \ac{its} scenario, consisting of cooperating vehicles (illustrated in Fig.~\ref{fig:scenario}). Each vehicle is equipped with a \ac{gnss}-type receiver to determine its absolute position and radar-type sensor to retrieve relative positions of mobile features present in the environment through \ac{v2f} measurements. 
Our goal is to develop a filter, which runs on an \ac{rsu}, to track the features and the states of all vehicles in every discrete time step $t$ through incorporation of all  measurements provided by the vehicle's on-board \ac{gnss} and \ac{v2f} sensors up until time step  $t$. %

\subsection{Vehicle and Feature Dynamics}
\label{sec:vehicleAndFeatureDynamics}

Vehicle state motion follows \ac{iid} Markovian processes, where the time-varying
vehicle state $\statex_{v,t}$ of each vehicle $v\in\mathcal{V}$
at time step $t$ is statistically modeled as
$p(\statex_{v,t}|\statex_{v,t-1})$.
Each feature $k\in\mathcal{K}$ with state $\statef_{k,t-1}$ survives to the next time step $t$ following an \ac{iid} Markovian process with survival probability $p_S(\statef_{k,t})$. The feature
state motion follows \ac{iid} Markovian processes and
is statistically modeled as $p(\statef_{k,t}|\statef_{k,t-1})$.
Note that
vehicle and feature state motion are independent.\footnote{Note that the proposed filter only predicts over small time-horizons in the order of a few tens of milliseconds. Then the assumption on vehicle and feature states evolving independently is reasonable, because there is very little interaction among them within the prediction horizon.}
In the following, we will drop the subscript indexing on states and measurements \ac{wrt} vehicle/feature/time whenever the context allows.

\subsection{Measurement Models}
\label{sec:measurment-models}\label{sec:V2FMeasurementModel}

At time step $t$, vehicle $v\in\mathcal{V}$ obtains two different
kinds of measurements: (i) measurements of the vehicle state $\statex$
\ac{wrt} the reference frame, i.e., \ac{gnss}-like measurements $\statey$
 modeled
by $p(\statey|\statex)$; and (ii)
measurements \ac{wrt} to features, i.e., from a radar-like (on-board)
\ac{v2f} sensor. Without loss of generality, we assume that the on-board sensors' state is equal to the vehicle state. %
Furthermore,  the
measurement-to-feature state correspondence  is not known and we assume that the standard \ac{mtt} measurement assumptions for point targets apply, i.e., each feature generates at most one measurement per measurement scan and sensor, and each measurement has a unique source (feature or clutter).

Let $\setZ$ be a set of \ac{v2f} measurements taken by a vehicle at a certain time. This set can be expressed as $ \setZ=\setZFA \uplus \setZD$, 
where $\setZFA$ denotes the set of false alarm measurements due to clutter (modeled by a \ac{ppp} with intensity $\lambda_c(\statez)$) and $\setZD$ denotes the set of feature measurements.
Let a \ac{v2f} measurement $\boldsymbol{z}\in \setZD$ 
be
obtained through the \ac{v2f} sensor with likelihood function $g(\boldsymbol{z}|\statex,\statef)$.
The likelihood for measurement set $\setZ=\{\statez_1,\ldots,\statez_m\}$ is defined as in \cite[Eqn.~13]{garcia2017poisson} by
\begin{align}
l(\setZ|\setF,\statex) &= e^{-\langle\lambda_c;1\rangle} \sum_{\setU \uplus \setF_1 \ldots \uplus \setF_m = \setF}
[1-p_D(\statex,\cdot)]^{\setU} \nonumber\\
&\times \prod_{i=1}^m\tilde{l}(\statez_i|\setF_i,\statex)
\label{eq:setLikelihood}
\end{align}
with
\begin{align}
\tilde{l}(\statez|\setF,\statex) &= \begin{cases}
p_D(\statex,\statef)g(\statez|\statex,\statef) & \setF=\{\statef\},\\
\lambda_c(\statez) & \setF=\emptyset,\\
0 & \vert\setF\vert>1.
\end{cases}
\end{align}
Here, the feature set has been decomposed into all possible sets $\setU\uplus\setF_1\ldots\uplus=\setF$, where the set $\setU$ represents unknown features and sets $\setF_i$ represent the origin of the $i$th measurement, which can either be a singleton containing the state of the feature that gave rise to the measurement or an empty set if the measurement is clutter, i.e., $|\setF_i|\leq1\forall i$. %
Above, $p_{\rm D}(\statex,\statef)$ denotes the probability of detection, which depends on the vehicle state $\statex$ as well as on the feature state $\statef$. For instance, a limited sensor \ac{fov} affects the probability of feature  detection based on the distance between vehicle and feature.
An alternative, but equivalent form to represent the likelihood \eqref{eq:setLikelihood}, is \cite[Eqns.~25, 26]{garcia2017poisson}
\begin{align}
l_o(\setZ|\setU,\setF_1,\ldots,\setF_n,\statex) &= \sum_{\setZ_1 \ldots \uplus \setZ_n \uplus \setZ^u = \setZ} l(\setZ^u|\setU,\statex)\nonumber\\
&\times \prod_{i=1}^n t(\setZ_i|\setF_i,\statex),
\label{eq:likelihoodlo}
\end{align}
where $|\setF_i|\leq 1$ for $i=1,\ldots,n$. Here, $l(\setZ^u|\setU,\statex)$ was defined in \eqref{eq:setLikelihood} and
\begin{align}
t(\setZ_i|\setF_i,\statex) &= \begin{cases}
p_D(\statex,\statef) g(\statez|\statex,\statef) & \setZ_i=\{\statez\}, \setF_i=\{\statef\},\\
1-p_D(\statex,\statef) & \setZ_i=\emptyset, \setF_i=\{\statef\},\\
1 & \setZ_i=\emptyset, \setF_i=\emptyset,\\
0 & \mathrm{otherwise}.
\end{cases}
\end{align}

\subsection{Communication}

We assume that every vehicle is able to communicate all obtained measurements (\ac{v2f} and \ac{gnss}) with the \ac{rsu} instantaneously
and without errors. This implies that at any time step $t$ the number of vehicles communicating with the \ac{rsu} can vary. 
The incorporation of a realistic \ac{v2i} channel model
and its performance impact is outside the scope of this work.


\section{Proposed Filter}
\label{sec:MBfiltering}
In  this section, we formulate the proposed Bayesian filter with uncertain vehicle (sensor) state. For complexity reasons, we aim for a recursive formulation with a factorized joint density over the vehicle and feature states. 
\subsection{Prior Joint State Density}
\label{sec:prior-density}
The vehicle state \ac{pdf} at time step $t-1$ is indicated by subscript '$-$', i.e., $p_-(\statex)$, the \ac{pdf} predicted to the current time step $t$ (before updating by a measurement) is indicated by subscript '$+$', i.e., $p_+(\statex)$, and the posterior \ac{pdf} is stated without subscript. Similar notation holds \ac{wrt} the feature \ac{rfs} density.
With a single vehicle, the prior joint vehicle-feature density is of form
\begin{align}
f_-(\statex,\setF)= p_-(\statex) f_-(\setF),
\label{eq:PMBpriorDensity}
\end{align}
where $p_-(\statex)$ is the prior \ac{pdf} on the vehicle state, and $f_-(\setF)$ is the prior \ac{pmbm} density.
The latter density can be written in terms of a \ac{ppp} density of unknown features $f_{-}^u(\setF^u)$, i.e., features which are hypothesized to exist but have never been detected \cite[Def.~I]{williams2015marginal}, and the prior \ac{mbm} \ac{rfs} density of detected features, as \cite[p. 484]{Mahler2007},\cite{williams2015marginal}
\begin{align}
f_{-}(\setF) & \propto \sum_{\setF^u \uplus \setF_1 \uplus \ldots \uplus \setF_n = \setF} f_{-}^u(\setF^u) \sum_{j\in\mathbb{J}} \prod_{i\in\mathbb{I}_j} w_{ji}f_{ji-}(\setF_i),
\label{eq:priorFeatureDensity}	
\end{align}
where the $w_{ji}$ and $f_{ji-}(\cdot)$ are the weight and Bernoulli density of potentially detected feature $i$ under global hypothesis $j$. Note that the weight of a global hypothesis $j$ is proportional to the single hypothesis weights by $\prod_{i\in\mathbb{I}_j}w_{ji}$. 
In \eqref{eq:priorFeatureDensity}, the \ac{ppp} density of unknown features is
\begin{align}
f_{-}^u(\setF^u) & = e^{-\conv{D_{-}^{u}}{1}} \prod_{\statef\in\setF^u} D_{-}^u(\statef),
\label{eq:priorPMBundetectedFeature}
\end{align}
where $ D_{-}^u(\statef)$ is the intensity of unknown features.
\subsection{Prediction Step}
\label{sec:prediction-step}
Due to the independent mobility of vehicle and features, the predicted joint vehicle-feature density is
\begin{align}
	f_{+}(\statex,\setF) = p_{+}(\statex)f_{+}(\setF),
\label{eq:predSensorFeatureDensity}	
\end{align}
where $p_{+}(\statex)$ is given by the Chapman-Kolmogorov equation  $p_{+}(\statex) = \int p(\statex|\statex^\prime)p_-(\statex^\prime) \diff\statex^\prime$, 
where $p(\statex|\statex^\prime)$ is the state transition \ac{pdf} %
and $p_-(\statex^\prime)$ is the prior \ac{pdf} \cite{arulampalam2002tutorial}.
Similarly, the predicted feature state \ac{pmb} density
is calculated by \cite[Ch.~14.3]{Mahler2007}
\begin{align}
f_{+}(\setF) = \int f(\setF|\setF^\prime)f_-(\setF^\prime) \delta \setF^\prime, 
\label{eq:predFeatureDensity}
\end{align}
where $f(\setF|\setF^\prime)$ is the transition \ac{rfs} density, and $f_-(\setF^\prime)$ is the prior \ac{pmb} density.
It can be shown that \eqref{eq:predFeatureDensity} is a \ac{pmb} density %
where the predicted intensity of unknown features $D_{+}^{u}(\statef)$ 
is given by \cite[Thm.~I]{williams2015marginal}
\begin{align}
D_{+}^{u}(\statef)=D_{}^{b}(\statef)+\int p_{\mathrm{S}}(\statef')p(\statef|\statef')D_{-}^{u}(\statef')\mathrm{d}\statef'.
\label{eq:intensityOfUndetectedFeatures}
\end{align}
Here, the (known) birth intensity is denoted $D_{}^{b}(\statef)$,  $p(\statef|\statef')$ denotes the feature transition \ac{pdf}, and the feature survival probability is denoted $p_{\mathrm{S}}(\statef')$.
The Bernoulli components of the \ac{mbm} %
are updated as follows \cite[Eqn. (40)]{williams2015marginal}: 
\begin{align}
p_{ji+}^{}(\statef) & = \int p_\mathrm{S}(\statef^{\prime} )p(\statef|\statef^{\prime })p_{ji-}(\statef^{\prime }) \diff\statef^{\prime } \\
r_{ji+} & = r_{ji-} \int p_\mathrm{S}(\statef^{\prime}) p_{ji-}(\statef^{\prime })\diff \statef^{\prime }.
\label{eq:predictedProbOfExistence}
\end{align}
where $p_{ji-}(\statef^{\prime })$ denotes the prior \ac{pdf} of the $ji$-th Bernoulli component.

\subsection{Measurement Update Step}
\label{sec:measurement-update-step}
Updating the joint vehicle-feature density \eqref{eq:predSensorFeatureDensity} by any of the two types of different measurements, \ac{gnss} and \ac{v2f} measurements, involves the application of Bayes' theorem. In the following, we describe the update calculations using the different type of measurements.
\subsubsection{Update with vehicle state measurement}
Let $\statey$ be a measurement related to the vehicle state $\statex$ through $p(\statey | \statex)$. Given
a predicted vehicle-feature density \eqref{eq:predSensorFeatureDensity}, by Bayes' theorem the updated density is
\begin{align}
	f(\statex,\setF | \statey) 
	& = \frac{p(\statey | \statex) p_{+}(\statex)}{\int p(\statey | \statex^\prime) p_{+}(\statex^\prime)\diff \statex^\prime } f_{+}(\setF).
	\label{eq:jointSensorFeatureDensityGPS}
\end{align}
In other words, the vehicle state density is updated with the measurement $\statey$, the feature set density is unaffected by the update, and the independent form is retained (c.f.~\eqref{eq:PMBpriorDensity}).
\subsubsection{Update with cluttered set of feature measurements}
\label{sec:updateWithClutteredSetOfFeatureMeasurements}
Let $\setZ$ be the \ac{v2f} measurement subject to the model defined in Section~\ref{sec:V2FMeasurementModel}.
Furthermore, we assume the probability of detection to be state-independent, i.e., $p_\mathrm{D}(\statex,\statef)=p_\mathrm{D}$. 
With the help of Bayes' rule, the updated joint vehicle-feature density is
\begin{align}
f(\statex,\setF|\setZ)\propto p_+(\statex)f_+(\setF)l(\setZ|\statex,\setF).
\label{eq:jointPosterior}
\end{align}
Due to the dependency of the \ac{v2f} measurements on  $\setF$ and $\statex$, the updated posterior density $\eqref{eq:jointPosterior}$ cannot easily be decomposed into the independent form of the prior joint density \eqref{eq:PMBpriorDensity}. 
In order to enable joint vehicle-feature state tracking with: (i) low complexity; and (ii) standard \ac{mtt} frameworks such as, e.g., \cite{williams2015marginal}, we calculate \eqref{eq:jointPosterior} approximately. 
With the help of Lemma~\ref{sec:lemma1} the joint vehicle-feature density is approximated as
\begin{align}
f(\statex,\setF|\setZ) \approx p(\statex|\setZ)f(\setF|\setZ),
\label{eq:jointPosteriorApprox}
\end{align}
where 
\begin{align}
p(\statex|\setZ) &\propto \int p_+(\statex)f_+(\setF)l(\setZ|\statex,\setF) \delta\setF,
\label{eq:marginalVehiclePost}\\
f(\setF|\setZ) &\propto \int p_+(\statex)f_+(\setF)l(\setZ|\statex,\setF) \diff\statex.
\label{eq:marginalFeaturePost}
\end{align}
In this way, the independent structure of the vehicle and feature density \eqref{eq:PMBpriorDensity} is retained. 
Note that it has been observed in the \ac{slam} context that a factorization such as \eqref{eq:jointPosteriorApprox} can generate optimistic estimates about the state uncertainty \cite{durrant2006simultaneous,castellanos1997building}.

An alternative to the approximation \eqref{eq:jointPosteriorApprox} is to perform a Rao-Blackwellization as done in \ac{slam} \cite{durrant2006simultaneous,lee2013slam}. Although more accurate, the extension of the multi-vehicle scenario is not apparent and opponent to a low complexity implementation we seek.

Now, only computation of the marginal posteriors remains. This is derived in the next two subsections. For convenience, let
\begin{align}
\mathcal{X}_{\setZ}(\statez)=\begin{cases}
0 & \statez\not\in\setZ,\\
1 & \statez\in\setZ,
\end{cases}
\end{align}
and
\begin{align}
\delta_\emptyset(\setY)=\begin{cases}
0 & \setY\neq\emptyset,\\
1 & \setY=\emptyset.
\end{cases}
\end{align}

\subsection{Marginal Feature Set Posterior} \label{sec:featurePosterior}

The feature posterior $f(\setF|\setZ)$ of \eqref{eq:jointPosteriorApprox} is given by the following Theorem.
\begin{theorem}
The feature posterior for measurement set $\setZ=\{\statez_1,\ldots,\statez_m\}$ can be approximated by a \ac{pmbm} with density
\begin{align}
f(\setF|\setZ) &\appropto
\sum_{\setY  \uplus \setF_1 \ldots \uplus \setF_n \uplus \setX_1 \ldots \uplus \setX_m = \setF}
f^\mathrm{}(\setY) \sum_j \nonumber\\
&\times\sum_{\setZ_1 \ldots \uplus \setZ_n \uplus \setZ^u = \setZ} 
 \prod_{i=1}^m \left[\mathcal{X}_{\setZ^u}(\statez_i)\rho^\mathrm{}(\statez_i)f^\mathrm{}(\setX_i|\statez_i)\right.\nonumber\\
&\left.+(1-\mathcal{X}_{\setZ^u}(\statez_i))\delta_\emptyset(\setX_i)\right]\nonumber\\
&\times\prod_{i=1}^n w_{ji}\rho_{ji}(\setZ_i)f_{ji}^\mathrm{}(\setF_i|\setZ_i),
\label{eq:featurePosteriorApprox}
\end{align}
where $\appropto$ means approximately equal to. Here, $f^\mathrm{}(\setY)$ represents the PPP component, and the functional forms of the densities $f^\mathrm{}(\setX_i|\statez_i)$ and $f_{ji}^\mathrm{}(\setF_i|\setZ_i)$, as well as the constants $\rho^\mathrm{}(\statez_i)$, $\rho_{ji}(\setZ_i)$ are provided in Appendix \ref{sec:Thm3Proof}. 
\end{theorem}

\begin{IEEEproof}
See Appendix \ref{sec:Thm3Proof}. 
\end{IEEEproof}

The feature posterior \eqref{eq:featurePosteriorApprox} consists of undetected features ($\setY$), hypotheses for newly detected features ($\setX_i$) and for updating existing features ($\setF_i$). This structure is of the same form\footnote{For the linear Gaussian case, the vehicle state uncertainty is essentially mapped onto the \ac{v2f} measurement uncertainty, which can be seen as an increased time-varying measurement variance. } as for a known vehicle (sensor) state, as in \cite{garcia2017poisson}, and is thus amenable for using a standard \ac{pmbm} filter implementation  \cite{williams2015marginal}. 
The weights of newly detected features are considered in the product over $m$ components (i.e., one new feature per measurement). %
A previously detected feature has an updated weight $w_{ji} \rho_{ji}(\setZ_i)$ consisting of the previous single hypothesis weight and, depending on whether $\setZ_i$ contains a measurement, the hypotheses for a detection or a miss of the feature.

Note that \eqref{eq:featurePosteriorApprox} is only approximately $f(\setF|\setZ)$ due to the approximations performed in the derivations. Therefore, the weights are also only approximate weights for $f(\setF|\setZ)$.

\subsection{Marginal Vehicle Posterior} \label{sec:vehiclePosterior}

We now proceed with the vehicle state posterior, which is stated in the following Theorem.

\begin{theorem}
The vehicle state posterior for measurement set $\setZ$ is
\begin{align}
 p(\statex|\setZ)  &  \appropto  
\sum_j \sum_{\setZ_1 \ldots \uplus \setZ_n \uplus \setZ^u = \setZ}  \prod_{\statez_i \in \setZ^u}
  \mathcal{X}_{\setZ^u}(\statez_i)\rho(\statez_i) \nonumber\\
&    
\times\prod_{i=1}^n 
 w_{ji}  \rho_{ji}(\setZ_i) q_j(\statex|\setZ_1,\ldots,\setZ_n,\setZ^u)
\label{eq:sensorPosterior}
\end{align}
where $q_j(\statex|\setZ_1,\ldots,\setZ_n,\setZ^u)$ is a properly normalized density defined in Appendix \ref{sec:Thm4Proof}. 
\end{theorem}

\begin{IEEEproof}
See Appendix \ref{sec:Thm4Proof}. 
\end{IEEEproof}

From the vehicle posterior \eqref{eq:sensorPosterior}, we observe that the single hypothesis weights $\rho^\mathrm{}(\statez_i)$ and $w_{ji} \rho_{ji}(\setZ_i)$ of each term are the same as in the feature posterior \eqref{eq:featurePosteriorApprox}, only the order of integration over the vehicle state $\statex$ and the single feature state $\statef$ is exchanged. Hence, both posteriors use the same approximate single hypothesis weights. Furthermore, we have obtained that a global hypothesis is approximately proportional to the product of the approximate single hypothesis weights.

\subsection{Implementation Aspects}
Here, we discuss approximations for a practical implementation to perform the sensor state \ac{v2f} measurement update when the probability of existence of detected features is high. Furthermore, we discuss the approximation of the joint \ac{da} and the reduction of the feature posterior density to contain only a single global hypothesis. Using this approximation, the complexity of the proposed filter is briefly discussed.
\subsubsection{Certain Feature Information}

The spatial \ac{pdf} of the \ac{ppp} modeling the undetected features needs to cover the whole space where new features appear. Due to this, a newly detected feature does not provide certain information to update the vehicle state and may be neglected in the vehicle state update.
Furthermore, one can approximate the vehicle state update by considering only previously detected features with a high existing probability (c.f. Sec.~\ref{sec:background-RFSstateEstimation}).
When detection probability is high, \eqref{eq:sensorPosterior} can be approximated by
\begin{align}
 p(\statex|\setZ)  &  \appropto 
\sum_j \sum_{\setZ_1 \ldots \uplus \setZ_n \uplus \setZ^u = \setZ}  \label{eq:sensorPosteriorsimple}\\
&  \times\left( \prod_{i=1}^n  w_{ji} 
\rho_{ji}(\setZ_i) \right) q_j(\statex|\setZ_1,\ldots,\setZ_n,\setZ^u), \nonumber 
\end{align}
where 
\begin{align}
& q_j(\statex|\setZ_1,\ldots,\setZ_n,\setZ^u)  \nonumber \\
&\appropto  p_+(\statex) \prod_{i=1}^n (1-\delta_\emptyset(\setZ_i))\int g(\statez_i|\statef_i,\statex)p_{ji}(\statef_i) \diff \statef_i. 
\end{align}

\subsubsection{Marginal Association}
For the feature update, the global hypothesis weights are proportional to $\prod_{\statez_i \in \setZ^u}\rho(\statez_i)\prod_{i=1}^n 
 w_{ji}  \rho_{ji}(\setZ_i)$. For the vehicle state update, the global hypothesis weight is proportional to $\prod_{i=1}^n  w_{ji} 
 \rho_{ji}(\setZ_i) $). A global hypothesis over all features can be approximated by the product of marginal association weights using the \ac{fg} approach of \cite{williams2014approximate}. 

For the feature state update, the \ac{tombp} filter \cite{williams2015marginal} permits the reduction of the posterior \ac{pmbm} density to a \ac{pmb} density containing only a single global hypothesis (so that the summation over $j$ disappears in the next time step).

For the vehicle state update, each likelihood term is weighted \ac{wrt} the single hypothesis. From the feature state update, the obtained marginal data association weights for the feature update can be reused in the vehicle state update, since they use the same weights for the same hypotheses. 
The weighted likelihoods times the prior vehicle state is then approximately proportional to the vehicle state posterior conditioned on the measurement set $\setZ$.

\subsubsection{Complexity}
For Gaussian linear models, a \ac{gnss} measurement update (c.f.~\eqref{eq:jointSensorFeatureDensityGPS}) has same complexity as a Kalman filter update. Using the marginal association proposed by the \ac{tombp} filter, the update of the feature state density by \ac{v2f} measurements has comparable complexity as an update step of the \ac{tombp} filter with the added complexity of the marginalization over the sensor state (c.f. Appendix~\ref{sec:Thm3Proof}). For the vehicle state update with the \ac{v2f} measurements we have the same number of hypotheses and hence similar complexity.

\subsection{Multi-Vehicle Generalization of Proposed Filter\label{subsec:multitarget-multisensor}}

Up to this point, we discussed joint vehicle-feature state tracking using a single vehicle, where \ac{gnss} and \ac{v2f} measurements are incorporated. To achieve feature tracking as described in Section~\ref{sec:motivation}, where a \ac{gnss} and a \ac{v2f} sensor is mounted on each vehicle, we have to consider the multi-vehicle/multisensor case, where each vehicle is equipped with a \ac{gnss} and a \ac{v2f} sensor. Since \ac{gnss} measurements are straightforward to deal with (they can be applied prior to the \ac{v2f} measurements), we focus only on \ac{v2f} measurements, considering a two-vehicle case. There are different approaches to handle the multi-vehicle setting, where we highlight two of them next.

\subsubsection{Parallel Approach}

Given time-synchronized measurements from 2 vehicles (with measurements $\setZ_1$ collected by vehicle $1$  and $\setZ_2$ collected by vehicle $2$, respectively), the joint posterior can be approximated by 
\begin{align}
& f(\statex_1,\statex_2,\setF|\setZ_1, \setZ_2)  \nonumber\\
&\approx p(\statex_1,\statex_2|\setZ_1, \setZ_2)f(\setF|\setZ_1, \setZ_2) 
\label{eq:multiSensorParallel}
\end{align}
similar to the approximation \eqref{eq:jointPosteriorApprox}.
Note that the set of global hypotheses is now the Cartesian product of the individual vehicle's set of hypotheses. Several different approaches can be employed to tackle this \ac{da} problem in a tractable manner. For instance, by employing sequential sensor-by-sensor measurement updates (also called iterator-corrector method) \cite{Mahler2007}, or by partition of the measurement set into subsets associated with the Bernoulli components \cite{saucan2017multisensor}, or by performing variational inference \cite{williams2016multiple}, or by solving the \ac{da} in parallel on a sensor-by-sensor basis \cite{meyer2018}.
Note that in a system where sensors are spatially distributed (c.f. Sec.~\ref{sec:motivation}) synchronization between sensors is involved and a sequential measurement update procedure may be used instead.

\subsubsection{Sequential Approach}
Here, we employ the sequential measurement update strategy together with the \ac{tombp} algorithm (c.f. Section~\ref{sec:measurement-update-step}). Note that in a real system sensors among different vehicles are difficult to synchronize and then this approach can be used. 
The update is first performed with respect to the first vehicle (c.f. Sec.~\ref{sec:featurePosterior} and Sec.~\ref{sec:vehiclePosterior}): 
\begin{align}
& f(\statex_1,\setF|\setZ_1) \appropto  \int p_+(\statex_1)l(\setZ_1|\statex_1,\setF)f_+(\setF)\delta \setF \nonumber\\
& \times \int p_+(\statex_1)l(\setZ_1|\statex_1,\setF)f_+(\setF)\diff \statex_1\\
& = p(\statex_1|\setZ_1)f(\setF|\setZ_1). 
\end{align}
Then, the density $f(\setF|\setZ_1)$ is used as a prior when performing the update with respect to the second vehicle: 
\begin{align}
& f(\statex_2,\setF|\setZ_1,\setZ_2) \appropto  \int p_+(\statex_2)l(\setZ_2|\statex_2,\setF)f(\setF|\setZ_1)\delta \setF \nonumber\\
& \times \int p_+(\statex_2)l(\setZ_2|\statex_2,\setF)f(\setF|\setZ_1)\diff\statex_2\\
& = p(\statex_2|\setZ_1,\setZ_2)f(\setF |\setZ_1,\setZ_2). 
\end{align}
Using this method, subsequent vehicles will benefit from updated vehicle and feature information of preceding vehicles. Note that contrary to \eqref{eq:multiSensorParallel}, the vehicle states are approximated here as being independent of each other. 
In our application example (c.f. Section~\ref{sec:motivation}), this means that an update of the joint vehicle-feature density, with measurements from a well-localized vehicle (certain vehicle state), results in an improvement of feature tracking performance when prior information on the features
is low. An update of the joint vehicle-feature density, with measurements from a poorly localized vehicle (uncertain vehicle state), permits the reduction of the uncertainty of its own vehicle state when prior information on the features is high.

%

\section{Results With Synthetic Data}
\label{sec:results-synthetic}

\begin{figure}
	\centering
	\footnotesize
	\includegraphics{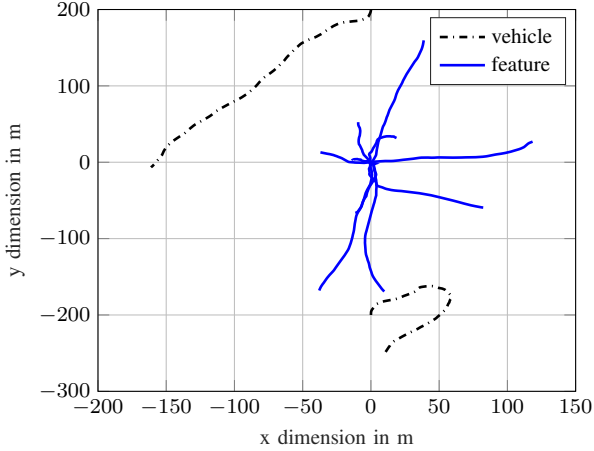}
	\caption{One realization of two vehicle and five feature trajectories for the experiment with synthetic data.
		\label{fig:trajectories}}
\end{figure}

We consider a scenario similar to the one outlined in Fig.~\ref{fig:scenario}, where we apply the proposed multisensor-multifeature state tracking filter presented in Section~\ref{sec:MBfiltering}.
\subsection{Setup}\label{sec:resultSetup}
The state of vehicle $v\in\mathcal{V}$ at time step $t$ is $\statex_{v,t}=[\boldsymbol{p}_{s,t}^{\mathsf{T}},\dot{\boldsymbol{p}}_{v,t}^{\mathrm{\mathsf{T}}}]^{\mathrm{\mathsf{T}}}$
with position $\boldsymbol{p}_{v,t}\in\mathbb{R}^{2}$ and velocity ${\dot{\boldsymbol{p}}}_{v,t}\in\mathbb{R}^{2}$.
We use a linear \ac{cv} model to model vehicle dynamics, where
\begin{equation}
\statex_{s,t}=\boldsymbol{A}\statex_{s,t-1}+\boldsymbol{w}_{s,t}\label{eq:SSMSensor}
\end{equation}
with state-transition matrix
\begin{align}
\boldsymbol{A}=\left[\begin{array}{cc}
1 & T_{s}\\
0 & 1
\end{array}\right]\otimes\boldsymbol{I}_{2},
\end{align}
where $T_{s}=0.5\;\mathrm{s}$ is the sampling time.
Above, $\boldsymbol{w}_{s,t}\sim\mathcal{N}(0,\boldsymbol{W})$ denotes the \ac{iid} process noise with
\begin{equation}
\boldsymbol{W}=r\left[\begin{array}{cc}
T_{s}^{3}/3 & T_{s}^{2}/2\\
T_{s}^{2}/2 & T_{s}
\end{array}\right]\otimes\boldsymbol{I}_{2}\label{eq:R},
\end{equation}
where $r=0.05~\mathrm{m}^{2}$. Above, $\otimes$ denotes
the Kronecker product. We set the initial state of vehicle 1 to $\statex_{1,0}=[0,200,0,-2]^\mathsf{T}$ and to $\statex_{2,0}=[0,-200,0,2]^\mathsf{T}$ for vehicle 2, respectively. 
The state of feature $k$ at time $t$, denoted $\statef_{k,t}\in\mathbb{R}^{4}$, is comprised of Cartesian
position and velocity, similar to the vehicle state $\statex_{v,t}$. There are at most five features present, if not noted otherwise. Furthermore,
feature dynamics follow the \ac{cv} model with the same parameters used
for the vehicles. To generate a challenging scenario for \ac{da}, we initialize
the feature states $\statef_{k,t}\sim\mathcal{N}(0,0.25\boldsymbol{I}_{4})$
at time step $t=175$ for all features $k\in\mathcal{K}$ and run the \ac{cv} model forward
and backward in time similar to \cite[Sec.~VI]{williams2015marginal}. The first feature enters the scene after $t=0$, the second after $t=20$ and so on. Once present, features stay alive for the remaining simulation time. A realization of vehicle and feature trajectories is shown in Fig.~\ref{fig:trajectories}.

For the \ac{gnss} measurements, we use the linear measurement model
\begin{equation}
\statey_{,t}=\boldsymbol{H}_{\mathrm{G}}\statex_t+\boldsymbol{r}_t,\label{eq:GNSSmeasurement}
\end{equation}
where the 
observation matrix is $\boldsymbol{H}_{\mathrm{G}}=\left[1\;0\right]\otimes\boldsymbol{I}_{2}$
and $\boldsymbol{r}_t\sim\mathcal{N}(0,\boldsymbol{R})$
denotes the measurement noise with
$\boldsymbol{R}_{}=\sigmaG\boldsymbol{I}_{2}.$ 
For vehicle $1,$ we assume it has low location uncertainty using \ac{rtkgps} with
$\sigmaG=5.76\cdot 10^{-4}\;\mathrm{m}^{2}$
and for vehicle $2$ high location uncertainty using \ac{spsgps} with $\sigmaG=12.96\;\mathrm{m}^{2}$, corresponding to a vehicle with high quality \ac{gnss} receiver and one with low quality \ac{gnss} receiver\footnote{According to \cite{hightsD52} and \cite{oxDatasheet}, the x/y position accuracy of the \ac{gnss} receiver RT3000 from OXTS is  $\sigmaG=12.96\;\mathrm{m}^{2}$ for \ac{spsgps},  $\sigmaG=2.0736\;\mathrm{m}^{2}$ for \ac{sbasgps},  $\sigmaG=0.9216\;\mathrm{m}^{2}$ for \ac{dgnss}, and  $\sigmaG=5.76\cdot 10^{-4}\;\mathrm{m}^{2}$ for \ac{rtkgps}.}. In the single sensor case, only vehicle $1$ is present, and in the multisensor case both vehicles are present, if not noted otherwise. 

For the \ac{v2f} measurements, we use the linear measurement model
\begin{equation}
\boldsymbol{z}_t=\boldsymbol{H}_{1}\statex_t+\boldsymbol{H}_{2}\statef_t+\boldsymbol{q}_t\label{eq:ZV2FModel}
\end{equation}
with $\boldsymbol{H}_{1}\triangleq\boldsymbol{H}_{\mathrm{G}}$ and $\boldsymbol{H}_{2}\triangleq-\boldsymbol{H}_{\mathrm{G}}$, and $\boldsymbol{q}_t\sim\mathcal{N}(0,\boldsymbol{Q})$
with measurement noise covariance matrix 
$\boldsymbol{Q}_{}=\sigmaVtoF\boldsymbol{I}_{2}$. 
Following \cite{williams2015marginal}, we set the initial unknown
feature intensity to $D^{u}_-(\statef_{k,t})=10\mathcal{N}(\boldsymbol{0},\boldsymbol{P}),$
where $\boldsymbol{P}=\mathrm{diag}([100^{2},100^{2},1,1]^{\mathsf{T}})$
to cover the ranges of interest of the feature state. The feature birth 
intensity is set to $D^{b}(\statef_{k,t})=0.05\mathcal{N}(\boldsymbol{0},\boldsymbol{P}),$
the average number of false alarms per scan to $\lambda_c=10,$
with uniform spatial distribution on $[-r_{\mathrm{max}},r_{\mathrm{max}}]$
with parameter $r_\mathrm{max}=500\;\mathrm{m}.$ Furthermore, the
probability of survival is $p_{\mathrm{S}}=0.7$ and the probability of detection is $p_{\mathrm{D}}(\statex_{v,t},\statef_{k,t})\triangleq p_{\mathrm{D}} = 0.9$. %
To assess feature tracking performance, we use the \ac{ospa} metric with cut-off parameter $c=20$ and order
$p=2$ \cite{schuhmacher2008consistent}. The filter tracking performance for the vehicle state at time step $t$
is assessed in terms of the position estimation error
$ e_t = \Vert \boldsymbol{p}_{t,\mathrm{true}} -  \hat{\boldsymbol{p}}_t \Vert_2$, 
where $\boldsymbol{p}_{t,\mathrm{true}}$ is the true vehicle position and $\hat{\boldsymbol{p}}_t$ the mean estimate of the filter.

We analyze the proposed \ac{mtt} filter \ac{wrt} feature tracking performance with sensor update from the tracked features to the sensor (proposed, sensor update), and without sensor update (proposed, no sensor update). For comparison, results are shown using the \ac{tombp} filter ignoring sensor state uncertainty \cite{williams2015marginal}, i.e., using the \ac{gnss} measurement as sensor state (\ac{tombp} I, no sensor update), and the \ac{tombp} filter ignoring sensor state uncertainty, but increasing the \ac{v2f} variance by the \ac{gnss} measurement variance (\ac{tombp} II, no sensor update), which is possible in the considered linear measurement scenario.
As a benchmark for vehicle localization performance, results from a centralized \ac{kf} are plotted as well, where measurement-to-feature \ac{da} is known and where the augmented state vector contains all vehicle and all feature states. This is denoted genie method. Furthermore, tracking performance using a local \ac{kf} is plotted. The local \ac{kf} performs filtering separately on the individual vehicle state using only \ac{gnss} measurements and does not estimate feature states. Note, the performance of the local \ac{kf} can be considered as the worst-case performance on vehicle state estimation, since \ac{v2f} measurements are not considered at all. 

\subsection{Discussion}
First, we discuss the impact of an uncertain vehicle state on feature tracking performance using a single vehicle and multiple features. After that, we consider the multisensor-multifeature case from Section~\ref{subsec:multitarget-multisensor}.%

\begin{figure}[t] 
	\centering
	\footnotesize
	\includegraphics{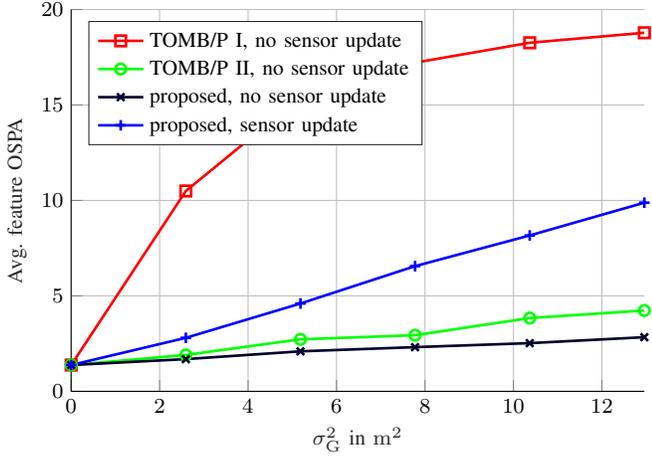} 
	\caption{Average feature OSPA for different values of GNSS measurement variance $\sigmaG$ using the proposed \ac{mtt} filter with sensor update from the tracked features to the sensor (proposed, sensor update) and without sensor update (proposed, no sensor update). For comparison, results using the \ac{tombp} filter ignoring sensor state uncertainty (\ac{tombp} I, no sensor update) and by increasing the \ac{v2f} noise variance by the \ac{gnss} variance (\ac{tombp} II, no sensor update) are shown. 
		\label{fig:F_OSPA_vs_GPSvar}}
\end{figure}

\begin{figure}[t] 
	\centering
	\footnotesize
	\includegraphics{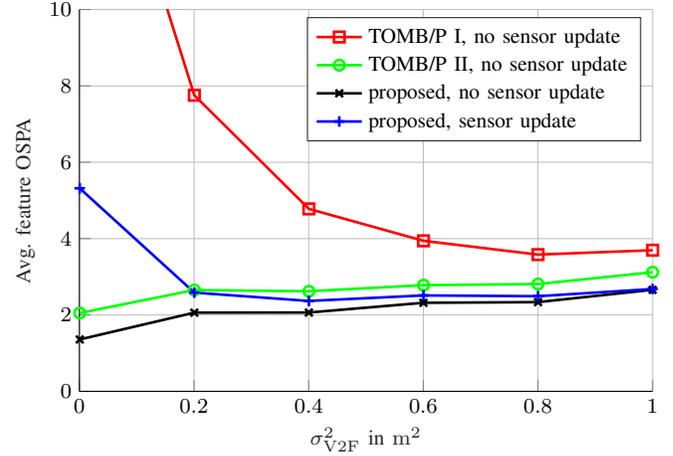}
	\caption{Average feature OSPA for different values of V2F measurement variance $\sigmaVtoF$. The \ac{gnss} noise variance is set to $\sigmaG=0.9216~\mathrm{m}^2$.
		\label{fig:F_OSPA_vs_V2Fvar}}
\end{figure}

\begin{figure}[t]
	\centering
	\footnotesize
	\includegraphics{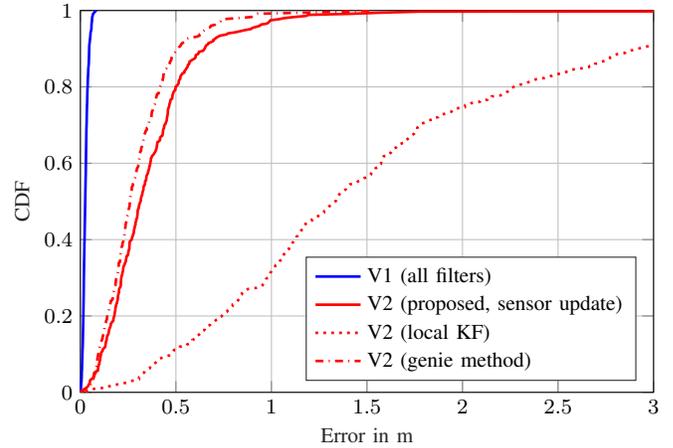}
	\caption{CDF plot of vehicle position estimation error. 
		\label{fig:CDF_vehicle}}
\end{figure}

\subsubsection{Impact of uncertain vehicle state on feature tracking performance}
For one simulation run, the features and vehicle trajectories are outlined in Fig.~\ref{fig:trajectories}. For the case the \ac{gnss} measurement variance is very small \ac{wrt} the \ac{v2f} measurement variance, e.g., $\sigmaG=5.76\cdot 10^{-4}~\mathrm{m}^2$ corresponding to \ac{rtkgps} and $\sigmaVtoF=0.42~\mathrm{m}^2$, the feature tracking performance of the proposed filter is comparable with the \ac{tombp} filter with known sensor state. Hence, we will not focus on this case and refer the reader to \cite{williams2015marginal} for performance results regarding the \ac{tombp} filter.

In Fig.~\ref{fig:F_OSPA_vs_GPSvar}, the feature state \ac{ospa} average of one simulation run of $351$ time steps averaged over $50$ \ac{mc} trajectory realizations is plotted for different values of \ac{gnss} measurement variance $\sigmaG$ for two variants of the proposed \ac{mtt} filter as well as with the \ac{tombp} filter. 
For all filter variants, the increase of \ac{gnss} measurement variance leads to an increased vehicle state uncertainty with the effect of an increase of the average feature \ac{ospa}. This \ac{ospa} increase consists of two components. First, an increased feature state estimation error due to a higher value of $\sigmaG$. Second, this results in features appearing in the filter spatially close together \ac{wrt} the feature state measurement uncertainty ($\sigmaG$ and $\sigmaVtoF$ together) for a longer period of time around time step $t=175$.
Hence, \ac{da} is more challenging with the effect of an increased feature \ac{ospa} in this regime. 

From the figure, we observe that not considering the present vehicle state uncertainty leads to the worst feature tracking performance (\ac{tombp} I). Incorporating this uncertainty by increasing the \ac{v2f} measurement variance improves feature tracking significantly. Modeling the present vehicle state uncertainty without artificially increasing the \ac{v2f} measurement variance, as proposed in this paper, leads to a slightly better feature tracking performance (proposed, no sensor update). For the case the vehicle state is updated by the \ac{v2f} measurement (proposed, sensor update), feature tracking performance deteriorates. The reason for this is that feature and vehicle state become correlated after the \ac{v2f} measurement update, which is not modeled by the proposed \ac{mtt} filter. This leads to the conclusion that the sensor should not use V2F measurements to update its own state if no other vehicles have updated the sensor state in the previous time step. With more vehicles, this effect of correlation will be diluted. 
Similar observations have been reported in \cite{frohle2018wcnc}.

In Fig.~\ref{fig:F_OSPA_vs_V2Fvar}, the average feature state \ac{ospa} is plotted for different values of \ac{v2f} measurement noise variance $\sigmaVtoF$ using the different filter variants. We observe that a higher \ac{v2f} noise variance leads to an increased average feature \ac{ospa} value except for the \ac{tombp} filter which ignores sensor state uncertainty (\ac{tombp} I). In the former, this reduction comes from the fact that sensor state uncertainty is absorbed from the increased \ac{v2f} noise variance. For all methods the single feature state estimation error increases and \ac{da} becomes more challenging. Since the sensor state uncertainty is negligible, all filter variants show the same performance. %

\subsubsection{Multisensor-multifeature tracking performance}
Here, we limit the discussion on the two vehicle (sensor) case since it is sufficient to analyze the effect of updating the vehicle state using \ac{v2f} measurements. %
In Fig.~\ref{fig:CDF_vehicle}, the \ac{cdf} of the vehicle position estimation error is plotted for the two vehicle scenario outlined in Sec.~\ref{sec:resultSetup}. 
We observe that for vehicle $1$, which has low \ac{gnss} measurement noise, all three filter methods deliver a similar performance. The reason for this is that, due to the high accuracy of \ac{gnss} measurements, not a lot of information (to improve the vehicle state) is provided from feature tracking.

Moving our focus to vehicle $2$, we observe that the error of the local \ac{kf} is much higher compared to the central \ac{kf}, which is caused by the high noise in the \ac{gnss} measurements. Due to the low estimation error of vehicle $1$'s state, there is relevant position information \textit{in the system}, which can be transfered from vehicle $1$ to vehicle $2$ via the features utilizing the \ac{v2f} measurements. 
In $80\%$ of all cases, the estimation error of vehicle $2$ is below $0.5~\mathrm{m}$ using the proposed filter, compared to $2.3~\mathrm{m}$ using the local \ac{kf}. 

Despite this great improvement of the proposed filter over the local \ac{kf}, it does not achieve the performance of the central \ac{kf} (genie method), where the error is around $0.4~\mathrm{m}$. The reason for the difference is that the central \ac{kf} has knowledge of the correct \ac{da}, knows the true number of features present, and  ignores clutter \ac{v2f} measurements. Furthermore, it tracks any present correlations between features and vehicles not modeled by the proposed filter. The proposed filter needs to infer the measurement-to-feature \ac{da}, estimate the number of features currently present, and needs to appropriately handle clutter in the \ac{v2f} measurement set $\setZ$.
%

\section{Results With Experimental Data}
\label{sec:results-experimental}
\subsection{Experiment Description}
Measurement data was recorded using the COPPLAR project test vehicle, a Volvo XC90 equipped with different kinds of automotive sensors. If not stated otherwise, the filter parameters are the same as described in Section~\ref{sec:resultSetup}.
The \ac{gnss} sensor is a high-precision Applanix POSLV, and the \ac{v2f} sensor is a onboard stereo vision camera from Autoliv looking in the car driving direction, which provides detections of objects \ac{wrt} the sensor coordinate frame. Measurements from both sensors arrive time-stamped, but are not synchronized. In order to obtain synchronized measurements, we linearly interpolated measurements from each sensor and sampled them at a lower rate of $T_s=0.1~\mathrm{s}$. \footnote{Note that this synchronization step is not needed, but it simplified analysis of the collected measurements \ac{wrt} filters' estimates.}

Since the coordinate frame of \ac{gnss} and \ac{v2f} sensor are different, we first mapped the \ac{gnss} measurements on the Cartesian coordinate system, and then used the heading measurements from the accurate \ac{gnss} sensor as ground truth to rotate and translate the \ac{v2f} measurements on the same coordinate system. This procedure allows to use the measurement models of Section~\ref{sec:measurment-models} without further modification.

Due to the absence of an exact ground truth in this dynamic measurement scenario, measurements were considered as ground truth and \ac{iid} zero-mean Gaussian measurement noise was artificially added to the \ac{gnss} measurements with $\sigmaG=0.9216~\mathrm{m}^2$ for vehicle $1$ corresponding to a \ac{dgnss} receiver and with $\sigmaG=12.95~\mathrm{m}^2$ for vehicle $2$ corresponding to a \ac{spsgps} receiver. Additionally, noise was added on the \ac{v2f} measurements with $\boldsymbol{Q}_t=0.42\boldsymbol{I}_2~\mathrm{m}^2$ which is the worst case performance of the stereo vision camera for object positioning.

In the scene (c.f. Fig.~\ref{fig:Copplar_trajectory}), there are two pedestrians (features) standing at an intersection and vehicle $1$ and $2$ are driving along perpendicular roads. Due to the limited sensor \ac{fov} of the camera, features become visible at approximately $50~\mathrm{m}$ distance. 
Since we had only one physical vehicle, we recorded first sensor data obtained by driving along one lane segment and afterwards from the perpendicular lane segment. The two vehicle driving scenario was then obtained by adjusting the time basis of one lane recording. Due to this hardware limitation, the proposed filter had to be run offline.
Note that additional results using the same hardware and the proposed filter have been reported in \cite{frohle2018ssp}.

\subsection{Discussion}

\begin{figure} 
	\centering
	\footnotesize
	\includegraphics{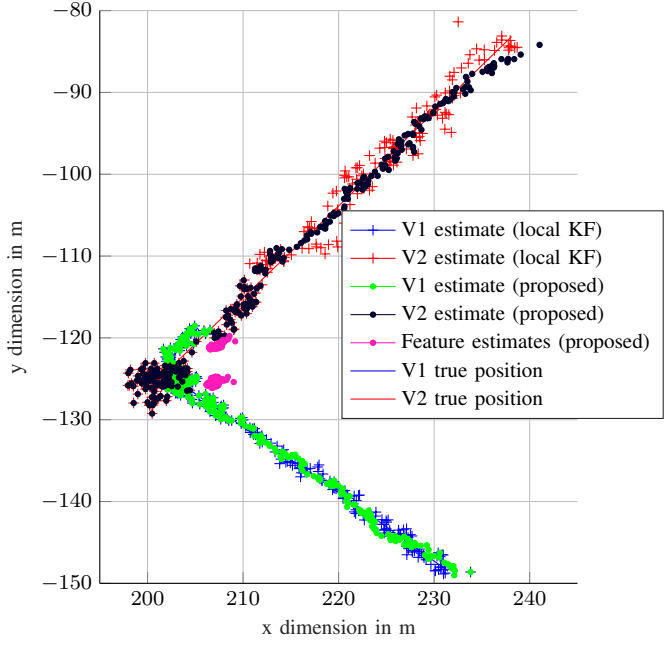}
	\caption{Estimated vehicle (V1, V2) and feature positions using a local \ac{kf} and with the proposed \ac{mtt} filter for one measurement realization. 
		\label{fig:Copplar_trajectory}}
\end{figure}

\begin{figure} 
	\centering
	\footnotesize
	\includegraphics{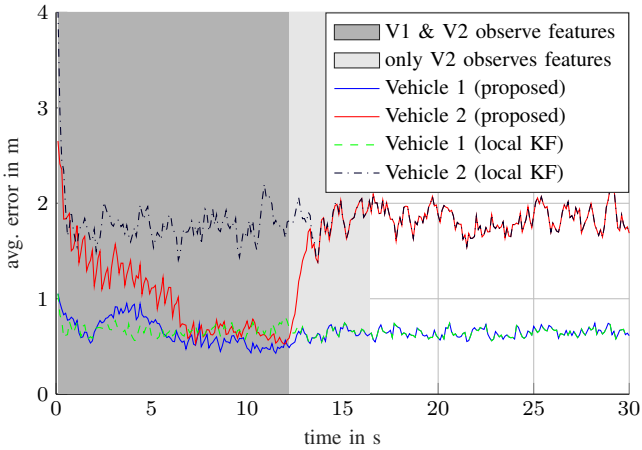}
	\caption{Average vehicle position estimation error is over time for experimental measurement data.
		\label{fig:Copplar_RMSE}}
\end{figure}

In Fig.~\ref{fig:Copplar_trajectory}, the trajectories of vehicle $1$ (V1) and vehicle $2$ (V2) are plotted together with the estimated feature and vehicle positions. Vehicle $1$ moving from the lower right upwards the left of the figure, uses a \ac{dgnss} receiver, and is therefore able to track the features quite accurately. In contrast, vehicle $2$, which uses a \ac{spsgps} receiver, cannot contribute much in accurate feature tracking, but its \ac{v2f} measurements allow to transfer position information to update the vehicle state. From the figure, we observe that the estimated vehicle track is much closer to the true trajectory compared to a local \ac{kf} using only \ac{gnss} measurements. As vehicle $2$ approaches the intersection, its performance deteriorates and achieves the same performance as the local \ac{kf}. The reason for this is that vehicle $1$ has already passed all features at this point in time and cannot provide tracking information on the features. Since we assume mobile features, no significant position information towards vehicle $2$ is provided after a few time steps. 

In Fig.~\ref{fig:Copplar_RMSE}, the vehicle position estimation error averaged over $50$ measurement realizations is plotted over time for this two vehicle experiment. Here, we can clearly observe how the performance of vehicle $2$ approaches the performance of vehicle $1$ when both vehicles observe the features (until approximately $12~\mathrm{s}$ time). It then deteriorates from $12~\mathrm{s}$ to $16~\mathrm{s}$ due to the absence of \ac{v2f} measurements from vehicle $1$. After approximately $16~\mathrm{s}$, vehicle $2$ has passed all features as well and performance of the proposed \ac{mtt} filter equals the performance of the local \ac{kf}. 
We can conclude that with the proposed multisensor \ac{mtt} filter, joint feature tracking provides relevant position information towards the vehicle's position allowing to improve the positioning quality for vehicles with different/varying position accuracies.
%

\section{Conclusions}
\label{sec:conclusions}

This paper presented a low-complexity Poisson multi-Bernoulli filter to jointly track multiple features (targets) as well as the state of multiple mobile sensors. This was enabled by an approximation minimizing the Kullback-Leibler divergence. 
Two different kinds of sensors providing observations of the sensor state itself as well as observations of features enable accurate feature and sensor state tracking.
The resulting filter incorporates the uncertain sensor state in the feature tracking task by marginalizing over the uncertain sensor state in the single feature state likelihood.
Information from multiple sensors is incorporated by asynchronous update steps, executed whenever
sensor measurements arrive at the central node. In doing so, data association is limited to a per-sensor basis.
Furthermore, in a multi-vehicle scenario with varying sensor qualities, an update of the uncertain sensor state is achieved by measurements of the sensor state and by means of feature tracking.
Simulation and experimental results showed that through the incorporation of measurements provided by different sensors, the feature tracking performance is superior to the \acf{tombp} filter which ignores sensor uncertainty, and comparable when its measurement noise is artificially increased. Furthermore, we observed in a multi-vehicle scenario that through joint vehicle-feature state tracking the vehicle state uncertainty can be significantly reduced compared to track the vehicle state alone. 
Applications of the proposed low-complexity filter involve cooperative vehicle driving scenarios when both information of the environment and the vehicle themselves are of interest. Furthermore, no direct vehicle-to-vehicle observations are needed, which makes it interesting for urban environments with applications such as extending the situational awareness beyond the visibility of the ego vehicle alone.

\appendices

\section{Proof of Theorem 3}\label{sec:Thm3Proof}
Plugging \eqref{eq:priorFeatureDensity} into  
 \eqref{eq:marginalFeaturePost} results in

\begin{align}
f(\setF|\setZ) &\propto  \int p_+(\statex) \sum_{\setU \uplus \setF_1 \ldots \uplus \setF_n = \setF} f_{+}^u(\setU) \sum_j\nonumber\\
&\times \prod_{i=1}^n w_{ji} f_{ji}(\setF_i)
l(\setZ|\setF,\statex)\diff\statex.
\end{align}
Now we add the constraint $|\setF_i|\leq1 \forall i$, since $f_{ji}(\setF_i)$ is Bernoulli and then replace the likelihood by \eqref{eq:likelihoodlo} leading to (c.f.~\cite{garcia2017poisson})
\begin{align}
f(\setF|\setZ)  \propto  \sum_{\setU \uplus \setF_1 \ldots \uplus \setF_n = \setF}
\sum_j\sum_{\setZ_1 \ldots \uplus \setZ_n \uplus \setZ^u = \setZ}   f_+^u(\setU) 
\label{eq:setFpost1}\\
\times
\underbrace{\int p_+(\statex) l(\setZ^u|\setU,\statex) \prod_{i=1}^n w_{ji} f_{ji}(\setF_i)t(\setZ_i|\setF_i,\statex)  \diff\statex}_{\approx q(\setZ^u|\setU)\prod_{i=1}^n w_{ji}q(\setZ_i|\setF_i)f_{ji}(\setF_i)},
\end{align}
where the approximation follows by invoking Lemma~\ref{sec:lemma1}, whereby the factors are found as follows: 
Using Lemma~\ref{sec:lemma1} again, we find
\begin{align}
q(\setZ^u|\setU) \approx  \sum_{\setY \uplus \setX_1 \ldots \uplus \setX_m = \setU}
e^{-\langle\lambda_c;1\rangle}
[1-p_D]^Y \prod_{i=1}^m \tilde{l}(\statez_i|\setX_i),
\end{align}
where 
\begin{align}
\tilde{l}(\statez_i|\setX_i)= \begin{cases}
p_D \int  g(\statez_i|\statex,\statef)p_+(\statex)\diff\statex & \setX_i=\{\statef\},\\
\lambda_c(\statez_i) & \setX_i=\emptyset,\\
0 & \mathrm{otherwise}.
\end{cases}
\end{align}
In a similar fashion, we obtain
\begin{align}
q(\setZ_i|\setF_i) &=
\begin{cases}
p_D \int  g(\statez|\statex,\statef) p_+(\statex)\diff\statex & \setZ_i=\{\statez\}, \setF_i=\{\statef\},\\
1-p_D & \setZ_i=\emptyset, \setF_i=\{\statef\},\\
1 & \setZ_i=\emptyset, \setF_i=\emptyset,\\
0 & \mathrm{otherwise}.
\end{cases}
\end{align}

Separating in \eqref{eq:setFpost1} all terms involving $\setU$, the update of the \ac{ppp}  and the newly detected features is given by  (c.f.~\cite[Eqns. 15 to 24]{garcia2017poisson})
\begin{align}
f^\mathrm{}(\setU|\setZ^u) & \propto f_+^u(\setU) q(\setZ^u|\setU)\nonumber\\
& \propto \sum_{\setY \uplus \setX_1 \ldots \uplus \setX_m = \setU} f^\mathrm{}(\setY) \prod_{i=1}^{m} f^\mathrm{}(\setX_i|\statez_i),
\label{eq:featureposteriorPPP}
\end{align}
where 
\begin{align}
& f^\mathrm{}(\setY) \propto [(1-p_D)D_{+}^{u}(\cdot)]^{\setY},\label{eq:postPPP}\\
& f^\mathrm{}(\setX_i|\statez_i) = 
\frac{\tilde{l}(\statez_i|\setX_i)f_+^u(\setX_i)}{e^{-\langle D_{+}^{u};1\rangle} \rho^\mathrm{}(\statez_i)}
\label{eq:qPostXizi}
\end{align}
and 
\begin{align}
\rho^\mathrm{}(\statez_i)&\triangleq \frac{\int \tilde{l}(\statez_i|\setX_i)f_+^u(\setX_i) \delta \setX_i}{e^{-\langle D_{+}^{u};1\rangle} \rho^\mathrm{}(\statez_i)} = 
\lambda_c(\statez_i)+e(\statez_i),\label{eq:rhoPostzi}\\
e(\statez_i) &= p_D\iint g(\statez_i|\statex,\statef)p_+(\statex)\diff\statex D_{+}^{u}(\statef)\diff\statef,
\end{align}
where we have defined $\rho^\mathrm{}(\statez_i)$ similar to \cite[Eqns. 19 to 21]{garcia2017poisson} by normalization over $e^{-\langle D_{+}^{u};1\rangle}$.

The updated Bernoulli components for the existing features are obtained by
\begin{align}
f_{ji}^\mathrm{}(\setF_i|\setZ_i) &= \frac{t(\setZ_i|\setF_i)f_{ji}(\setF_i)}{\rho_{ji}(\setZ_i)}\label{eq:qjiPostFiZi},
\end{align}
where 
\begin{align}
\rho_{ji}(\setZ) &= 
\begin{cases}
p_D r_{ji}\iint g(\statez|\statex,\statef)p_+(\statex)\diff\statex p_{ji}(\statef)\diff\statef & \setZ=\{\statez\},\\
1-p_D r_{ji} & \setZ=\emptyset
\end{cases}
\label{eq:rhojiZ}
\end{align}
and
\begin{align}
t(\setZ_i|\setF_i) &= \int t(\setZ_i|\setF_i,\statex)p_+(\statex)\diff\statex.
\end{align}

\subsection*{Parameters of the Bernoulli Components}
From the derivation above, it is possible to find the parameters of all the Bernoulli components modeling newly detected features and existing features. 
\subsubsection*{Newly Detected Features}
The posterior for a newly detected feature, denoted $f^\mathrm{}(\setX_i|\statez_i)$ (c.f.~\eqref{eq:qPostXizi}), has Bernoulli components with parameters
\begin{align}
r^\mathrm{}(\statez_i) &= \frac{e(\statez_i)}{\rho^\mathrm{}(\statez_i)}, \label{eq:rPostzi}\\
p^\mathrm{}(\statef|\statez_i) &= \frac{p_D D_{+}^{u}(\statef) \int g(\statez_i|\statex,\statef)p_+(\statex)\diff\statex }{e(\statez_i)}.
\label{eq:pPostzi}
\end{align}

\subsubsection*{Existing Features}
The posterior of an existing Bernoulli component, denoted $f_{ji}^\mathrm{}(\setF_i|\setZ_i)$ (c.f.~\eqref{eq:qjiPostFiZi}), has updated parameters
\begin{align}
r_{ji}(\setZ_i)&=
\begin{cases}
1 & \setZ_i=\{\statez\},\\
\frac{r_{ji}(1-p_D)}{1-p_D r_{ji}} & \setZ_i=\{\emptyset\},
\end{cases}\label{eq:rjiPost}\\
p_{ji}(\statef | \setZ_i) &\propto
\begin{cases}
(1-p_D) p_{ji}(\statef) & \setZ_i=\{\emptyset\},\\
p_D p_{ji}(\statef)\int g(\statez|\statex,\statef)p_+(\statex)\diff\statex  & \setZ_i=\{\statez\}.
\end{cases}
\label{eq:pjiPost}
\end{align}

\section{Proof of Theorem 4}\label{sec:Thm4Proof}

Plugging \eqref{eq:priorFeatureDensity} into  
\eqref{eq:marginalFeaturePost} results in
\begin{align}
p(\statex|\setZ) & \propto p_+(\statex) \int  \sum_{\setU \uplus \setF_1 \ldots \uplus \setF_n = \setF} f_{+}^u(\setU) \sum_j\nonumber\\
&\times
\prod_{i=1}^n w_{ji} f_{ji}(\setF_i)
l(\setZ|\setF,\statex)\delta\setF.
\end{align}
Similar to the feature posterior, we add the constraint $|\setF_i|\leq1 \forall i$, since $f_{ji}(\setF_i)$ is Bernoulli, replace the likelihood by \eqref{eq:likelihoodlo}, and make use of Lemma~\ref{sec:lemma2} resulting in
\begin{align}
& p(\statex|\setZ)    \propto
p_+(\statex) \int  \sum_{\setU \uplus \setF_1 \ldots \uplus \setF_n = \setF: |\setF_i|\leq1 \forall i}\sum_j\nonumber\\
& \times  \sum_{\setZ_1 \ldots \uplus \setZ_n \uplus \setZ^u = \setZ} 
f_{+}^u(\setU)\prod_{i=1}^n w_{ji} f_{ji}(\setF_i)
l(\setZ|\setF,\statex)\delta \setF\\
& =
 \sum_j\sum_{\setZ_1 \ldots \uplus \setZ_n \uplus \setZ^u = \setZ}   p_+(\statex) q(\setZ^u|\statex) \prod_{i=1}^n w_{ji} q_{ji}(\setZ_i|\statex) , \label{eq:statepost1}
\end{align}
where
\begin{align}
& q(\setZ^u|\statex) \propto 
{\left[\lambda_c(\statez_i)+p_D\int g(\statez_i|\statex,\statef) D_{+}^{u}(\statef)\diff\statef\right]}
\end{align}
and
\begin{align}
q_{ji}(\setZ_i|\statex) = {\int t(\setZ_i|\setF_i,\statex)f_{ji}(\setF_i) \delta\setF_i}.
\end{align}
To obtain the global hypothesis weight, we express \eqref{eq:statepost1} as a mixture of vehicle state distributions. The weights are given by
\begin{align}
\int q(\setZ^u|\statex)\prod_{i=1}^n w_{ji} q_{ji}(\setZ_i|\statex) p_+(\statex)\diff\statex \nonumber\\
\overset{}{\approx} \int q(\setZ^u|\statex)p_+(\statex)\diff\statex \prod_{i=1}^n  w_{ji}  \int q_{ji}(\setZ_i|\statex) p_+(\statex)\diff\statex,
\label{eq:weightApproxSensor}
\end{align}
where the approximation is due to Lemma~\ref{sec:lemma1}. 
We now look at the factors individually. Applying again Lemma~\ref{sec:lemma1}, the first factor can be approximated by
\begin{align}
\int q(\setZ^u|\statex)p_+(\statex)\diff\statex 
\appropto
\prod_{\statez_i \in \setZ^u}
\rho(\statez_i),
\end{align}
where $\rho(\statez_i)$ was defined in \eqref{eq:rhoPostzi}, with the only difference that the order of integration is exchanged.

For each factor in the product term in \eqref{eq:weightApproxSensor}, we get
\begin{align}
 \int q_{ji}(\setZ_i|\statex) p_+(\statex)\diff\statex = \rho_{ji}(\setZ_i),
\end{align}
where $\rho_{ji}(\setZ_i)$ was defined in \eqref{eq:rhojiZ}. 
Substitution of the terms above into \eqref{eq:statepost1} yields the vehicle state posterior \eqref{eq:sensorPosterior}, where
\begin{align}
q_j(\statex|\setZ_1,\ldots,\setZ_n,\setZ^u) &\nonumber\\
= \frac{p_+(\statex) q(\setZ^u|\statex) \prod_{i=1}^n  q_{ji}(\setZ_i|\statex)}{\int q(\setZ^u|\statex)\prod_{i=1}^n q_{ji}(\setZ_i|\statex) p_+(\statex)\diff\statex}.
\end{align}
\bibliographystyle{IEEEtran}
\bibliography{bib}
%
%
%

\vspace{-1.cm}
\begin{IEEEbiography}[{\includegraphics[width=1in,height=1.25in,clip,keepaspectratio]{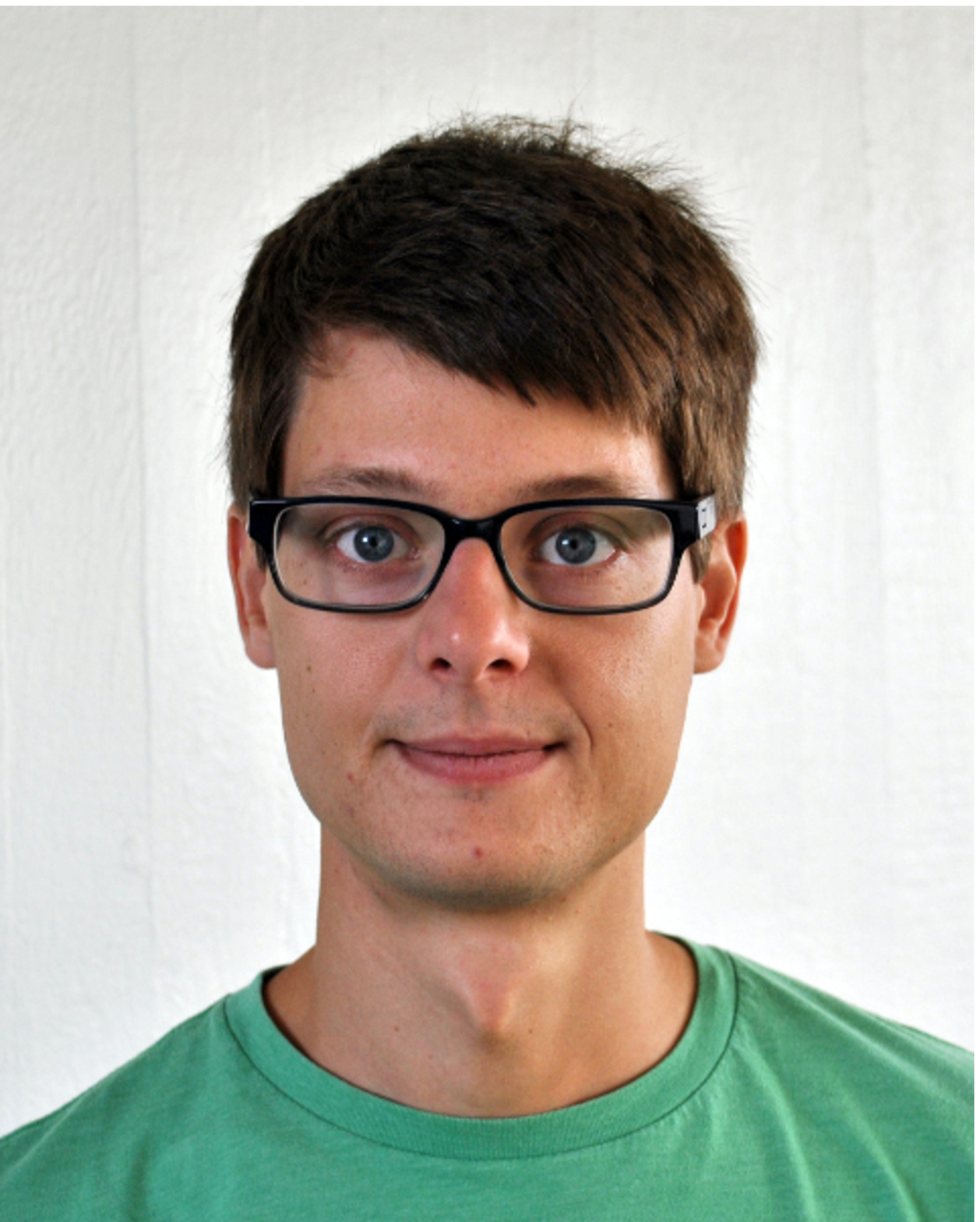}}]{Markus~Fr\"ohle}
	(S'11) received the B.Sc. and M.Sc. degrees in Telematics from Graz University of Technology, Graz, Austria, in 2009 and 2012, respectively. From 2012 to 2013, he was a Research Assistant with the Signal Processing and Speech Communication Laboratory, Graz University of Technology. Since 2013 he is working towards the Ph.D. degree in electrical engineering with the Department of Electrical Engineering, Chalmers University of Technology, Gothenburg, Sweden. His current research interests include signal processing for wireless multi-agent systems, and localization and tracking.
\end{IEEEbiography}
\vspace{-1.cm}
\begin{IEEEbiography}[{\includegraphics[width=1in,height=1.25in,clip,keepaspectratio]{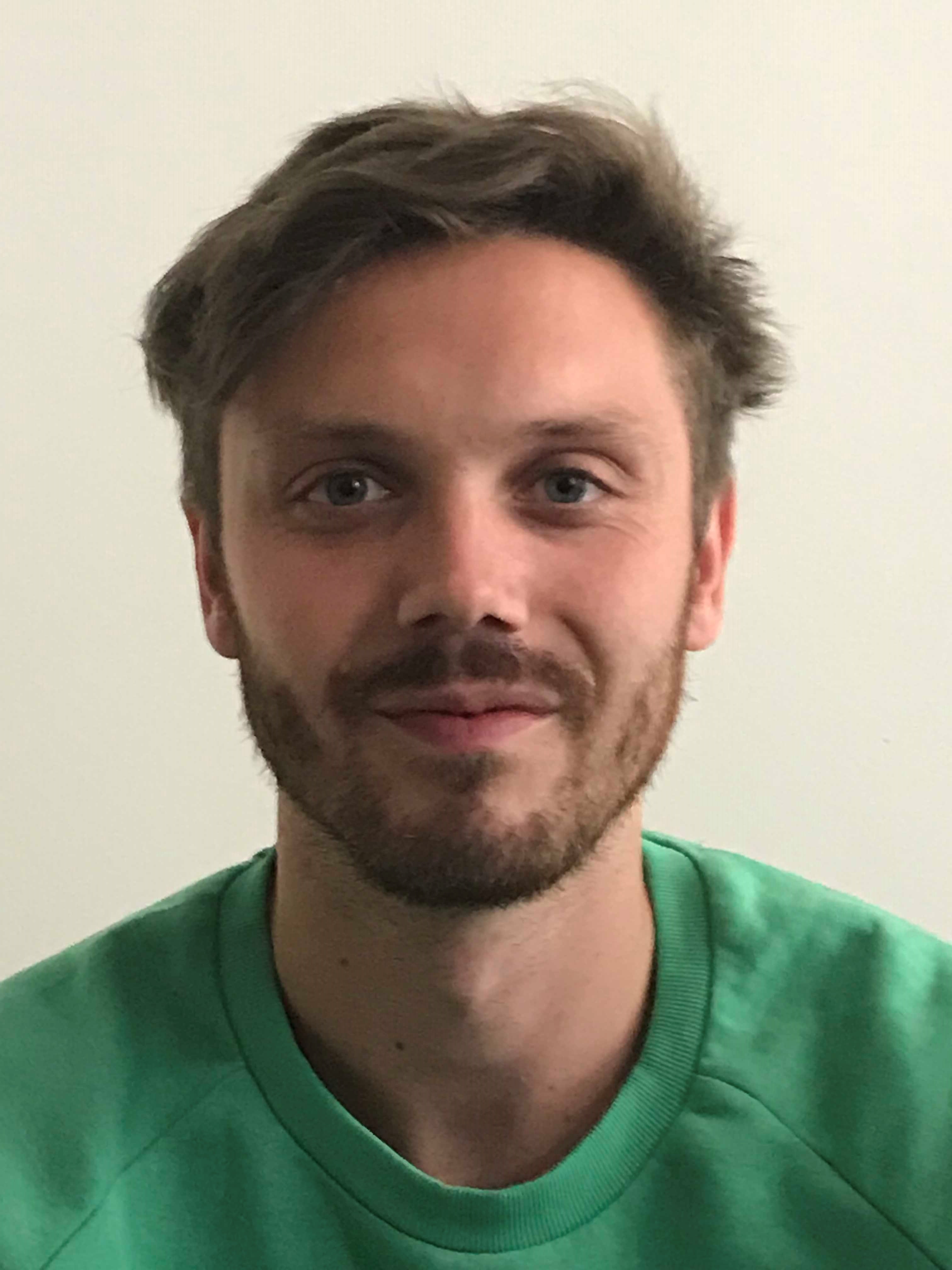}}]{Christopher~Lindberg} received the B.Sc. degree in electrical engineering, M.Sc. degree in engineering mathematics and computational science, and degree of Licentiate of Engineering in electrical engineering from Chalmers University of Technology in 2010, 2012, and 2015, respectively. Since 2012 he is working towards the Ph.D. degree in the Communication Systems group at the Department of Electrical Engineering, Chalmers University of Technology, Gothenburg, Sweden. Currently he is working with sensor fusion for AD/ADAS at DENSO Sweden AB. His research interests involve cooperative distributed processing in networks based on consensus algorithms and belief propagation. 
\end{IEEEbiography}
\vspace{-1.cm}
\begin{IEEEbiography}[{\includegraphics[width=1in,height=1.25in,clip,keepaspectratio]{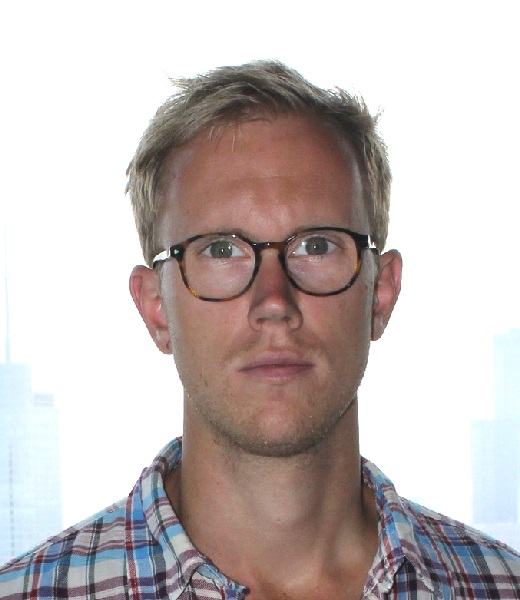}}]{Karl Granstr\"{o}m} (M'08) is a postdoctoral research fellow at the Department of Signals and Systems, Chalmers University of Technology, Gothenburg, Sweden. He received the MSc degree in Applied Physics and Electrical Engineering in May 2008, and the PhD degree in Automatic Control in November 2012, both from Link\"{o}ping University, Sweden. He previously held postdoctoral positions at the Department of Electrical and Computer Engineering at University of Connecticut, USA, from September 2014 to August 2015, and at the Department of Electrical Engineering of Link\"{o}ping University from December 2012 to August 2014. His research interests include estimation theory, multiple model estimation, sensor fusion and target tracking, especially for extended targets. He received paper awards at the Fusion 2011 and Fusion 2012 conferences. He has organised several workshops and tutorials on the topic Multiple Extended Target Tracking and Sensor Fusion.%
\end{IEEEbiography}
\vspace{-1.cm}
\begin{IEEEbiography}[{\includegraphics[width=1in,height=1.25in,clip,keepaspectratio]{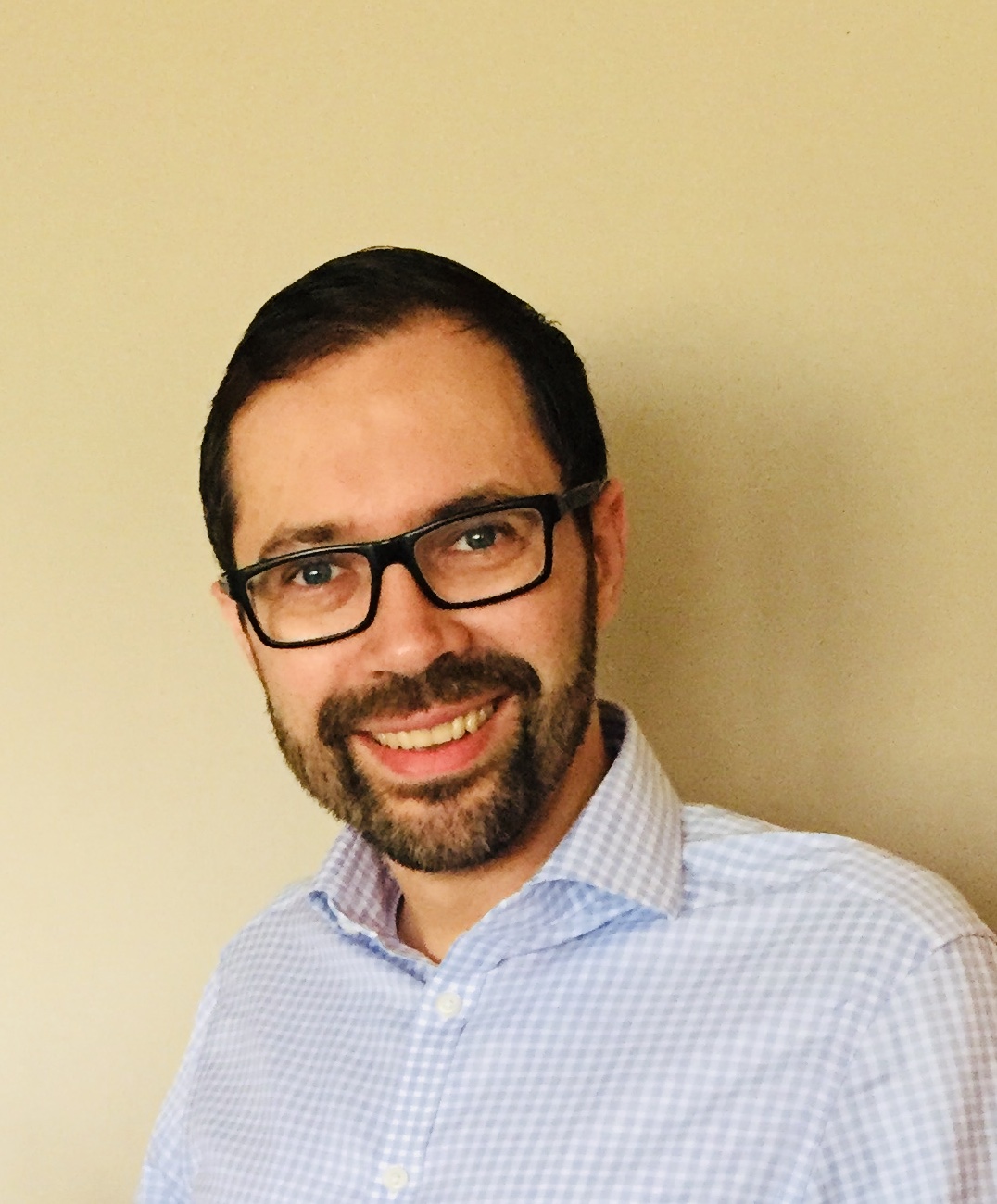}}]{Henk~Wymeersch}
	(S'01, M'05) obtained the Ph.D. degree in Electrical Engineering/Applied Sciences in 2005 from Ghent University, Belgium. He is currently a Professor of Communication Systems with the Department of Electrical Engineering at Chalmers University of Technology, Sweden. Prior to joining Chalmers, he was a postdoctoral researcher from 2005 until 2009 with the Laboratory for Information and Decision Systems at the Massachusetts Institute of Technology. Prof. Wymeersch served as Associate Editor for IEEE Communication Letters (2009-2013), IEEE Transactions on Wireless Communications (since 2013), and IEEE Transactions on Communications (2016-2018). His current research interests include cooperative systems and intelligent transportation. 
\end{IEEEbiography}

\end{document}